\documentclass[pre,nofootinbib,superscriptaddress,showpacs,showkeys]{revtex4}
\usepackage{graphicx}
\usepackage{amsmath}

\newcommand{\rcite}[1]{Ref.~\onlinecite{#1}}
\newcommand{\rcites}[1]{Refs.~\onlinecite{#1}}
\newcommand{\eq}[1]{Eq.~(\ref{#1})}

\newcommand{\epl}{{EPL}}

\newcommand{\br}{{\bf r}}
\newcommand{\bF}{{\bf F}}
\newcommand{\brp}{{\bf r}_\parallel}
\newcommand{\en}{{\bf e}_n}
\newcommand{\et}{{\bf e}_t}
\newcommand{\ez}{{\bf e}_z}
\newcommand{\ex}{{\bf e}_x}
\newcommand{\ey}{{\bf e}_y}
\newcommand{\er}{{\bf e}_r}

\newcommand{\bn}{{\bf n}}
\newcommand{\nablap}{\nabla_\parallel}
\newcommand{\stress}{\mathsf{T}}
\newcommand{\rdrop}{R_\mathrm{d}}
\newcommand{\vepsf}{\boldsymbol{\varepsilon}_\mathrm{el}}
\newcommand{\vepse}{\boldsymbol{\varepsilon}_\mathrm{ext}}
\newcommand{\vepsp}{\boldsymbol{\varepsilon}_\Pi}
\newcommand{\upd}{{\rm d}}

\begin{document}

\title{Force balance of particles trapped at fluid interfaces}

\author{Alvaro Dom{\'\i}nguez}
\email{dominguez@us.es}
\affiliation{F\'\i sica Te\'orica, Universidad de Sevilla, Apdo.~1065, 
  E--41080 Sevilla, Spain}
\author{Martin Oettel}
\affiliation{Institut f\"ur Physik, Universit{\"a}t Mainz, 
  D-55029 Mainz, Germany}
\author{S.\ Dietrich}
\affiliation{Max--Planck--Institut f\"ur Metallforschung, 
  Heisenbergstr.\ 3, D-70569 Stuttgart, Germany}
\affiliation{Institut f\"ur Theoretische und Angewandte Physik, 
  Universit{\"a}t Stuttgart, Pfaffenwaldring 57,
  D--70569 Stuttgart, Germany}

\date{Mar 21, 2008}


\begin{abstract}
  We study the effective forces acting between colloidal particles trapped at a fluid
  interface which itself is exposed to a pressure field.
  To this end we apply what we call the ``force approach'', which relies solely on
  the condition of mechanical equilibrium and turns to be in a certain sense
  less restrictive than the more frequently used ``energy approach'',
  which is based on the minimization of a free energy functional.
  The goals are (i) to elucidate the advantages and disadvantages of the force
  approach as compared to the energy approach, and (ii) to disentangle
  which features of the interfacial deformation and of the
  capillary--induced forces between the particles follow from
  the gross feature of
  mechanical equilibrium alone, as opposed to features which depend on details
  of, e.g., the interaction of the interface with the
  particles or the boundaries of the system.
  First, we derive a general stress--tensor formulation of the forces
  at the interface. On that basis we work out a useful analogy with 2D 
  electrostatics in the particular case of small deformations of the interface 
  relative to its flat configuration. We
  apply this analogy in order to compute the asymptotic decay of
  the effective force between particles
  trapped at a fluid interface,
  extending the validity of previous results
  and revealing the advantages and limitations of the force approach 
  compared to the energy approach.
  It follows the application of the force approach to the case of
  deformations of a non--flat interface.
  In this context we first compute
  the deformation of a spherical droplet due to the
  electric field of a charged particle trapped at its surface and conclude
  that the interparticle capillary force is unlikely to explain certain recent
  experimental observations within such a configuration.
  We finally discuss the application of our approach to a generally curved
  interface and show as an illustrative example that a nonspherical
  particle deposited on an interface forming a minimal surface is pulled to 
  regions of larger curvature.
\end{abstract}

\maketitle

\section{Introduction}
\label{sec:intro}

Experimental evidence has been accumulated that electrically charged,
$\mu$m--sized colloidal particles trapped at fluid interfaces can
exhibit long--ranged attraction despite their like charges
\cite{GhEa97,RGI97,SDJ00,QMMH01,NBHD02,TAKK03,FMMH04,CTNF05}. The
mechanisms leading to this attraction are not yet fully understood.
An attraction mediated by the interface deformation was proposed
\cite{NBHD02},
in analogy to the capillary force due to the weight of large floating
particles \cite{Nico49,CHW81}. However, for the particles sizes used
in the aforementioned experiments gravity is irrelevant. Instead,
one is led to invoke
electrostatic forces which act on the interface. This feature has
triggered investigations of capillary deformation and
capillary--induced forces beyond the well studied case of an
interface simply under the effects of gravity and surface tension
\cite{MeAi03,FoWu04,ODD05a,ODD05b,WuFo05,DOD05,ODD06,DKB06,DOD06a}.
These studies have relied almost exclusively on what we shall call the {\em
  ``energy approach''}, which is based on the minimization of a free
energy functional obtained as a parametric function of the positions
of the particles by integrating out the interfacial degrees of
freedom, leading to a ``potential of mean force''.
This functional has to include the contribution by the
interface itself, by the particles, and by the boundaries of the system.
Moreover, due to technical challenges the theoretical implementation of this approach 
is de facto restricted
to the regime of small interfacial deformations.

In the following, as an alternative we investigate
the {\em ``force approach''} which follows by directly applying the
condition of mechanical equilibrium.
Our analysis is based on the pressure field $\Pi(\br)$ (generated, e.g., by
electrostatic forces) acting on the interface between two fluid
phases. In general, the condition that an arbitrary piece ${\cal S}$
of this interface is in mechanical equilibrium reads (see, e.g., \rcite{Sege77})
\begin{equation}
  \label{eq:equilibrium}
  \int_{\cal S} \upd A \; \en \Pi + 
  \gamma \oint_{\partial{\cal S}} \upd\ell \; \et \times \en = 
  {\bf 0} ,
\end{equation}
where $\en$ is the local unit vector normal to the interface, $\et$ is the
unit vector tangent to the boundary $\partial {\cal S}$ (oriented such
that $\et \times \en$ points towards the exterior of ${\cal S}$),
$\upd A$ is the element of the interfacial area, $\upd\ell$ is the element
of the arclength along the contour $\partial {\cal S}$, and $\gamma$ is
the (spatially homogeneous) surface tension of the interface. 
In \eq{eq:equilibrium}, the first term is the so-called {\em bulk}
force exerted on the piece ${\cal S}$ by the pressure $\Pi$ and the
second one is the {\em line} force exerted on the contour and
generated by the surface tension (also called capillary force).
This equation is the starting point for the subsequent calculations.

The force approach allows us to obtain new results, to derive
previous ones more easily than within the energy approach,
and to gain additional insight. 
This approach was employed in \rcite{KPDI93} for the special case that gravity is
the only relevant force and it was shown to give the same results as
the energy approach if the deformations with respect to a flat interface are
small everywhere. For an arbitrary pressure field acting on the interface, in
\rcite{DOD05} we applied the force approach in order to obtain the deformation
of an otherwise flat interface far from the particles generating it.
In the following we further illustrate the force approach:
In Sec.~\ref{sec:stress} we first express the force exerted by the
interface in terms of a stress--tensor formulation, extending a recent
result \cite{MDG05a} to the most general case $\Pi(\br)
\neq$~constant. This formulation also allows us to establish a useful
analogy between two--dimensional electrostatics and the description of
small capillary deformations of a flat interface. Since this analogy has been
already employed by several authors in a more or less explicit manner,
here we present a thorough discussion addressing not only the issue of
interfacial deformations, but also that of boundary conditions and of the
capillary forces.
In Sec.~\ref{sec:force} we exploit the stress--tensor formulation and the electrostatic
analogy in order to study the interface--mediated effective force between
colloidal particles trapped at a fluid interface and to provide a 
detailed comparison with
the corresponding results obtained within the energy approach. 
In Sec.~\ref{sec:nonflat} we compute
the deformation of a spherical droplet due to the presence of a
charged particle at its surface, generalizing a corresponding result obtained in
\rcite{DOD05} for a flat interface and correcting certain claims in
the literature. Finally we discuss the more general case that the
unperturbed interface is curved.
Sec.~\ref{sec:end} provides a summary and an outlook.

\section{Stress--tensor formulation and electrostatic analogy}
\label{sec:stress}

The capillary force exerted by the interface (second term in
\eq{eq:equilibrium}) can be rewritten as
\begin{equation}
  \label{eq:defT}
  \gamma \oint_{\partial{\cal S}} \upd\ell \; \et \times \en = 
  \mbox{} - \oint_{\partial{\cal S}} 
  \upd\ell \; (\et \times \en) \cdot \stress , 
\end{equation}
which serves to define the stress tensor $\stress (\br) := \mbox{} -
\gamma \mathsf{1}(\br)$, where $\mathsf{1}(\br \in {\cal S})$ is the
2D identity tensor on the tangent plane of ${\cal S}$ at each point
$\br$.
In these terms, the condition of mechanical equilibrium (\eq{eq:equilibrium})
takes the form
\begin{equation}
  \label{eq:equilibrium2}
  \int_{\cal S} \upd A \; \en \Pi =
  \oint_{\partial{\cal S}} 
  \upd\ell \; (\et \times \en) \cdot \stress . 
\end{equation}
We recall that $\et \times \en$ is a vector tangent to the surface
${\cal S}$ but normal to the contour $\partial{\cal S}$ and pointing
outwards. This allows one to reinterpret \eq{eq:equilibrium2} as the definition of the stress tensor $\stress$, the flux of which
through a closed boundary is the bulk force (first term
in \eq{eq:equilibrium}) acting on the piece of interface enclosed by
that boundary.
In dyadic notation one has (summation over repeated indices is
implied)
\begin{equation}
  \stress =  \mbox{} - \gamma \mathsf{1} = 
  \mbox{} - \gamma \, g^{ab} \, {\bf e}_a \, {\bf e}_b 
  \quad \Rightarrow \quad
  T^{ab} =  \mbox{} - \gamma \, g^{ab} ,
\end{equation}
where $\{{\bf e}_1, {\bf e}_2\}$ is a local basis, at each
point tangent to the surface ${\cal S}$, and $g_{ab} = {\bf e}_a \cdot {\bf
  e}_b$ is the induced metric
($g^{ab}$ are the contravariant components of this tensor).
In this form, we have the same stress
tensor as the one derived in \rcite{MDG05a} using methods of
differential geometry within the energy approach and for the
restricted case $\Pi(\br)=\rm{constant}$. 
We remark, however, that \eq{eq:equilibrium2} holds for an arbitrary
pressure field $\Pi(\br)$.

An analogy with 2D electrostatics emerges by considering 
small deformations relative to a {\it flat} interface\footnote{More
  precisely, the small quantity is the spatial gradient of the
  deformation (see Eqs.~(\ref{eq:quasiflat}, \ref{eq:Eanalogy})).}
corresponding to the generic experimental set-up, 
i.e., a situation like the one in Fig.~\ref{fig:ref} but with, e.g., a charged
particle \cite{GhEa97,RGI97,ACNP00,QMMH01,TAKK03,FMMH04,CTNF05}, a
nonspherical particle \cite{SDJ00,LYP06}, or a droplet at a nematic interface \cite{SCLN04}, so that the interface is
deformed by an electric field, by a nonplanar contact line,
or by the elastic stress in the nematic phase,
respectively. To this end, we identify the flat interface with the
$XY$--plane, so that any point of the deformed interface can be expressed as $\br
=\brp + \ez u(\brp)$ with $\brp := x\ex + y\ey$, and
(the subscript $_\parallel$ will be used to denote quantities
evaluated at and operators acting in the reference, flat interface)
\begin{subequations}
  \label{eq:quasiflat}
\begin{eqnarray}
  \en \, \upd A & = & 
  \left(\frac{\partial \br}{\partial x} \times 
    \frac{\partial \br}{\partial y} \right) 
  \upd y \upd x = 
  \left[ \ez - \nablap u \right] \upd A_\parallel , \\
  & & \nonumber \\
  \et \, \upd\ell & = & 
  (\upd\boldsymbol{\ell}_\parallel \cdot \nabla_\parallel) \br = 
  \upd\boldsymbol{\ell}_\parallel + 
  \ez (\upd\boldsymbol{\ell}_\parallel \cdot \nablap u) ,
\end{eqnarray}
\end{subequations}
where $\upd \boldsymbol{\ell}_\parallel = \ex \upd x + \ey \upd y$,
$\upd A_\parallel = \upd x \upd y$, and $\nablap = \ex
(\partial/\partial x) + \ey (\partial/\partial y)$. 
We denote the projection of any piece of interface ${\cal S}$ onto the
$XY$--plane as ${\cal S}_\parallel$, and introduce the unit vector
$\bn$ in the $XY$--plane which is normal to the contour $\partial{\cal
  S}_\parallel$ and points outwards. With this, we expand
\eq{eq:equilibrium2} in terms of the deformation $u(\brp)$; to lowest order 
the component in the direction of $\ez$ is linear in $u(\brp)$,
\begin{subequations}
  \label{eq:Eanalogy}
  \begin{equation}
    \label{eq:force_z}
    \int_{{\cal S}_\parallel} \upd A_\parallel \; \Pi =
    \mbox{} - \gamma \oint_{\partial{\cal S}_\parallel} 
    \upd\ell_\parallel \; \bn \cdot \nablap u .
\end{equation}
The local version of this equality is the linearized
Young--Laplace equation:
\begin{equation}
  \label{eq:linYL}
  \gamma \nablap^2 u = - \Pi .   
\end{equation}
To lowest order the components of \eq{eq:equilibrium2} in the
$XY$--plane are quadratic in the deformation:
\begin{equation}
  \label{eq:force_xy}
  \mbox{} - 
  \int_{{\cal S}_\parallel} \upd A_\parallel \; \Pi \; \nablap u = 
  \oint_{\partial{\cal S}_\parallel} 
  \upd\ell_\parallel \; \bn \cdot \stress_{\parallel} , 
\end{equation}
where
\begin{equation}
  \label{eq:Tpar}
  \stress_{\parallel}\ := \gamma \left[ 
    (\nablap u) (\nablap u) - \frac{1}{2} |\nablap u|^2 
    \; \mathsf{1}_\parallel \right] 
\end{equation}
\end{subequations}
is a stress tensor defined in the $XY$--plane. We remark that
\eq{eq:force_xy} also implies \eq{eq:linYL} upon applying Gauss'
theorem, demonstrating consistency.

The form of \eq{eq:linYL} allows us to identify $u$ with an
electrostatic potential 
(and $-\nablap u$ with an electric field), 
$\Pi$ with a charge density 
(``capillary charge'' \cite{KDD01}), and $\gamma$ with a 
permittivity\footnote{There is also a dual magnetostatic
  interpretation in terms of magnetic fields created by currents along
  the $Z$--direction; the correspondences are $\ez u(\brp)
  \leftrightarrow {\bf A}(\brp)$, $\nablap \times (\ez u)
  \leftrightarrow {\bf B}(\brp)$, $\gamma \leftrightarrow 1/\mu_0$,
  $\ez \Pi(\brp) \leftrightarrow {\bf j}(\brp)$.}. The boundary
conditions usually imposed on
$u$ at a contour ${\cal C}_\parallel$ have a close electrostatic
analogy, too (see Fig.~\ref{fig:bc} for the notation):

\begin{figure}
  \centering{\includegraphics[width=.85\textwidth]{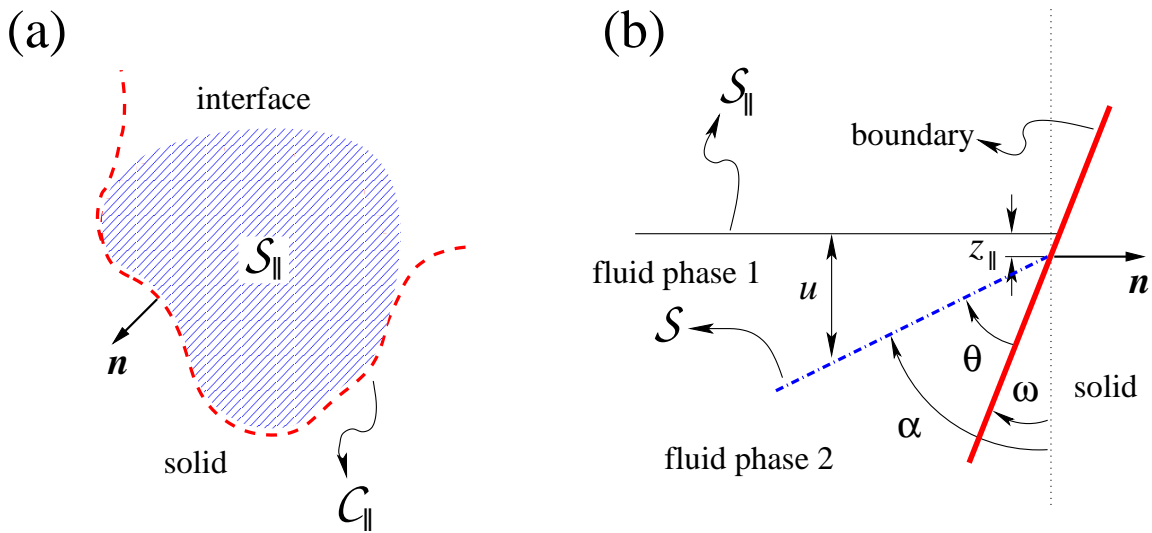}}
  \caption{Generic view of the three-phase contact line of an
    interface with a solid boundary. (a) Top view: projection ${\cal
      S}_\parallel$ onto the $XY$-plane of the interface (dashed area)
    near the projected contact line ${\cal C}_\parallel$ (dashed
    line). The unit vector $\bn$ is normal to the projected contact
    line and is directed towards the exterior of ${\cal S}_\parallel$.
    (b) Side view of the same configuration within a vertical plane
    containing $\bn$. The thin horizontal line is the section of
    ${\cal S}_\parallel$ (i.e., the flat, unperturbed interface),
    while the dash--dotted line is the section of the actual interface
    ${\cal S}$, which can be approximated locally by the tangent
    forming an angle $\theta$ (contact angle) with the solid boundary.
    The latter is the full thick line, which looks locally like a
    straight line, in general inclined by an angle $\omega$ with
    respect to the vertical direction (dotted line). The height of the interface
    at contact entering into the boundary
    condition~(\ref{eq:pinnedbc}) is $z_\parallel:=z(\brp)$.  The
    boundary condition~(\ref{eq:anglebc}) expresses the slope of the
    interface at contact with $\alpha := \theta + \omega$.}
  \label{fig:bc}
\end{figure}

\noindent
(i) The potential is given at ${\cal C}_\parallel$. $\Leftrightarrow$
The interface is pinned at the contour at a height $z(\brp)$:
\begin{equation}
  \label{eq:pinnedbc}
  u(\brp) = z (\brp), \qquad \brp \in {\cal C}_\parallel .
\end{equation}

\noindent
(ii) The normal component of the electric field is given at ${\cal
  C}_\parallel$.  $\Leftrightarrow$ The contact angle $\theta(\brp)$
is specified at each point of the contour: 
\begin{equation}
  \label{eq:anglebc}
  (\bn \cdot \nablap) u(\brp) = \cot \alpha(\brp), \qquad 
  \brp \in {\cal C}_\parallel ,
\end{equation}
where the angle $\alpha(\brp)$, defined in Fig.~\ref{fig:bc}, must be
close to $\pi/2$ for reasons of consistency with the approximation of
small deformations.

This means that the equivalence with the corresponding electrostatic
analogy is exact concerning the relationship between the deformation
field $u(\brp)$ and its sources (i.e., the pressure field $\Pi$ and
the boundary conditions).
The analogy carries over almost exactly, too, to the elastic forces in the $XY$--plane
(``lateral capillary forces'') arising from interfacial deformation:
according to Eqs.~(\ref{eq:force_xy},
\ref{eq:Tpar}), $\stress_\parallel$ corresponds to Maxwell's stress
tensor, the flux of which through a closed boundary gives the electric
force acting on the enclosed charge. However, the force related to the
deformation is actually the interfacial stress, which is {\rm minus}
the flux of this tensor (see \eq{eq:defT}). Therefore, the
electrostatic analogy is valid up to a reversal of the forces and the
peculiarity arises that ``capillary charges'' will attract~(repel)
each other if they have equal~(different) sign. The origin of this
peculiarity is that \eq{eq:equilibrium2}, which  in
the spirit of electrostatics can be reinterpreted as a definition of $\stress$, 
is actually
a relationship between bulk and capillary forces as two physically different 
kinds of forces. The actual connection of $\stress$ with a
force is \eq{eq:defT}.

We note that the electrostatic analogy holds wherever the
deformation of the interface is small, even if there are other regions
of the interface where this is not true. Such ``nonlinear patches''
can be surrounded by contours where the deformation is small, so that
the values of the field $u$ and of its derivatives at these contours
play the role of a boundary condition for the ``linear patches''.
This means that the ``nonlinear patches'' are replaced by a
distribution of virtual ``capillary charges'' inside the corresponding
regions. In particular, there is a simple physical meaning associated
with the total ``capillary charge'' and the total ``capillary
dipole''. The ``capillary charge'' $Q$ of a piece of interface bounded
by a contour ${\cal C}$ is given by Gauss' theorem solely in terms of
the value of the deformation at the contour (see \eq{eq:force_z}):
\begin{equation}
  \label{eq:gauss}
  Q = \mbox{} - \gamma \oint_{{\cal C}_{\parallel}} 
  \upd\ell_\parallel \; \bn \cdot \nablap u .
\end{equation}
The right hand side of this equation is minus the capillary force
exerted on the piece of interface in the $Z$--direction. This implies
that in terms of the bulk force ${\bf F}_{\rm bulk}$ and by virtue of
the condition of mechanical equilibrium one has
\begin{equation}
  \label{eq:mono_general}
  Q = \ez \cdot {\bf F}_{\rm bulk} .
\end{equation}
This holds even if the deformation in the bulk (i.e., inside the
contour ${\cal C}$) is not small. In the same manner, it can be shown
(see Appendix~\ref{app:dipole})
that the total ``capillary dipole'' ${\bf P}$ (with respect to the
origin of the coordinate system) and the torque ${\bf M}_{\rm bulk}$
(with respect to the same origin) exerted by
the bulk force are related according to
\begin{equation}
  \label{eq:dipole}
  {\bf P} = \ez \times {\bf M}_{\rm bulk} .
\end{equation}

The electrostatic analogy provides a transparent visualization of
small interfacial deformations and ensuing forces in terms of a 2D
electrostatic problem. We note that in \rcite{Paun98} such a kind of
analogy is also established, but between the capillary interaction and
the 3D electrostatic problem (DLVO theory, see, e.g., \rcite{KlLa03})
for that very same 3D geometrical set-up.
As a side remark we note that if gravity (or the disjoining pressure due to a substrate)
is relevant, it contributes a pressure field which depends explicitly
on the deformation field: $\Pi_\mathrm{grav}= - \gamma u / \lambda^2$
with the capillary length $\lambda$ \cite{DOD06a}. This replaces
\eq{eq:linYL} by a different equation which is formally equivalent to
the field equation of the Debye--H\"uckel theory for dilute
electrolytes (see, e.g., \rcite{KlLa03}) with $\lambda$ playing the role of the Debye length. This
suggests that extending the electrostatic analogy to this case
is possible, but this task is beyond the scope of the present effort.

\section{Effective interparticle forces}
\label{sec:force}
 
\begin{figure}
  \centering{\includegraphics[width=.65\textwidth]{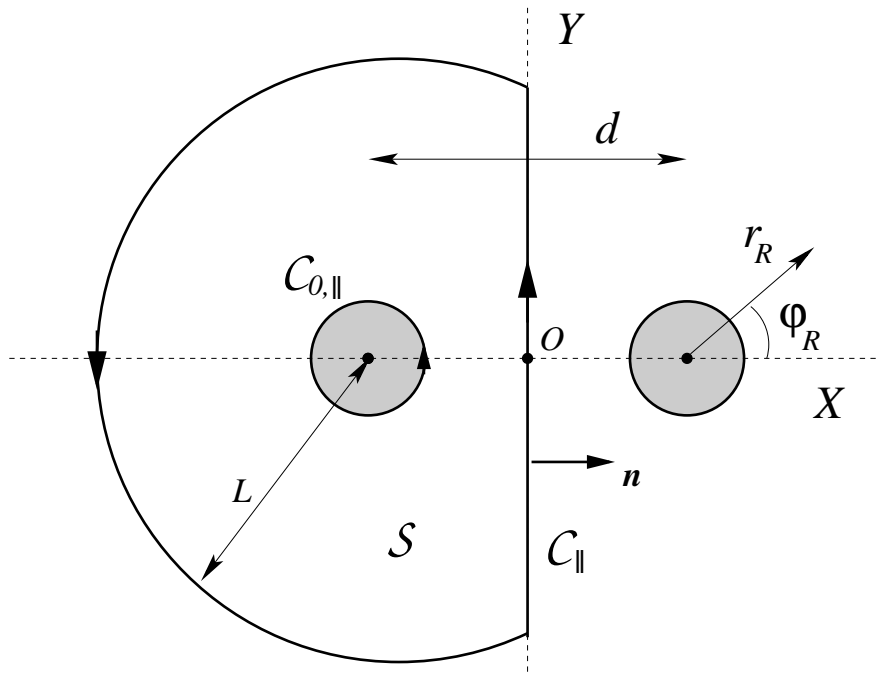}}
  \caption{Configuration of two particles (top view) a lateral
    distance $d$ apart, trapped at an asymptotically flat interface.
    The left half ${\cal S}$ of the full interface is bounded by the
    particle--interface contact line ${\cal C}_0$ (generically
    noncircular), the oriented projection of which onto the
    $XY$--plane is ${\cal C}_{0,\parallel}$, and a contour ${\cal C}$,
    the oriented projection ${\cal C}_\parallel$ of which consists of
    a piece of the $Y$--axis and a circle of radius $L\to\infty$. The
    unit vector $\bn$ is normal to the contour and directed towards
    the exterior of ${\cal S}$. The origin of the coordinate system is
    located at the midpoint $O$, while the coordinates $r_R$,
    $\varphi_R$ parametrize the plane with respect the center of the
    right particle.}
  \label{fig:2coll}
\end{figure}

The capillary deformation may give rise to an effective attraction
between two identical particles trapped at the interface which could
explain certain corresponding experimental observations. We compute this force using
the 
electrostatic analogy derived
above. The study of this issue will also clarify
the general relationship between the force and the energy approaches as well as
the respective advantages and disadvantages.
We consider the equilibrium configuration of two 
particles at
an asymptotically flat interface and fixed to be a distance $d$ apart (see Fig.~\ref{fig:2coll}).
The total capillary force $\bF_\mathrm{total}$ acting on the left half (which includes the piece
${\cal S}$ {\em and} the enclosed particle) is given by \eq{eq:defT}:
\begin{equation}
  \label{eq:Ftotal}
  \bF_\mathrm{total} = 
  \mbox{} - 
  \oint_{{\cal C}} 
  \upd\ell \; (\et \times \en) \cdot \stress .
\end{equation}
If the particle separation $d$ is large enough, we can assume that the
deformation of the interface {\em at} (but not necessarily inside) the
contour ${\cal C}$
is small, so that the lateral force
$\bF_\parallel$ (= component of $\bF_\mathrm{total}$ in the
$XY$--plane = $\bF_\mathrm{total}-(\bF_\mathrm{total}\cdot\ez)\ez$) 
is (see \eq{eq:force_xy})
\begin{equation}
  \label{eq:Flateral}
  \bF_\parallel \approx  \mbox{} - \oint_{{\cal C}_\parallel} 
  \upd\ell_\parallel \; \bn \cdot \stress_\parallel . 
\end{equation}
In the limit $L\to\infty$, the contribution from the circular part of ${\cal C}_\parallel$ vanishes and only the knowledge of the deformations at the 
straight midline part of ${\cal C}_\parallel$ is required, for
which the electrostatic analogy will hold provided the interfacial
deformations are small there.
Thus, one can try to estimate the ``electric potential'' $u$ at the
midline as $d\to\infty$
by a multipole expansion. 
The deformation field $u_R$ created by the right
half plane 
behaves asymptotically like (see Appendix~\ref{app:multipolar})
\begin{equation}
  \label{eq:multipole}
  u_R(\br_R) \sim \frac{Q_0}{2\pi\gamma} \ln \frac{\zeta}{r_R}
  + \frac{1}{2\pi\gamma} \sum_{s=1}^\nu \frac{Q_s \,{\rm e}^{-i s \varphi_R} + Q_s^* \,{\rm e}^{i s \varphi_R}}{2 s (r_R)^s} 
  + \Delta u_R .
\end{equation}
Here, ${\bf r}_R = \brp - \ex d/2$ is the position of a point relative
to the center of the right particle (see Fig.~\ref{fig:2coll}), $\zeta$ is a
fixed length determined by the distant boundary conditions, which set the
zero point (undeformed interface) of the ``electric potential'' $u_R$, and $Q_s$ is given in terms
of the $2^s$--pole of ``capillary
charge'' associated with the particle and the surrounding interfacial deformation; if this is small everywhere, it is
\begin{equation}
  \label{eq:charge}
   Q_s = \int \upd A_R \; 
   (r_R)^s \, {\rm e}^{i s \varphi_R} \, \Pi_{\rm (single)} (\br_R) 
   + \tilde{Q}_s^{\rm (single)},
\end{equation}
with $\tilde{Q}_s^{\rm (single)}$ the corresponding charge associated to the
particle (which is defined by the multipole
expansion (see \eq{eq:multexpansion}) applied to the interfacial
deformation at the contact line).
The multipole expansion is based on the implicit assumption that the main
source of the deformation field is concentrated at or near the
particles. For that reason $Q_s$ is given by the ``capillary charge''
distribution for $d=\infty$, i.e., for the single--particle
configuration.
The asymptotically subdominant term $\Delta u_R$ in \eq{eq:multipole}
accounts for the corrections to this assumption, i.e.,
``polarization'' effects by the second particle and the fact that,
even in the single--particle configuration,
the pressure field $\Pi$ is expected to decay smoothly 
asymptotically far from the particles rather
than dropping exactly to zero beyond some distance: if $\Pi
(r_\parallel\to\infty) \sim r_\parallel^{-n}$ (actually $n=6$ in
realistic models in the case of electric stresses
\cite{Hurd85,ACNP00,DaKr06a,DFO08}, and $n=8$ in the case of nematic stresses
\cite{ODTD07}), there is the bound $\nu<n-2$ 
(see the sum in \eq{eq:multipole} and Appendix~\ref{app:multipolar}).

In general the leading term in the multipole expansion is determined by the
``capillary monopole'' $Q_0$. 
Thus asymptotically for $d\to\infty$, $\bF_\parallel$ will be given by
the ``electric force'' between two monopoles:
\begin{equation}
  \label{eq:Flog}
  \bF_\parallel \approx \frac{Q_0^2}{2\pi\gamma d} \ex ,
\end{equation}
with the reversed sign as explained in the previous Section, so that
it describes an attraction. Since $Q_0$ is given in terms of the net
bulk force according to \eq{eq:mono_general}, the $1/d$-dependence
only arises if there is an external field acting on the system. For
example, \eq{eq:Flog} corresponds to the ``flotation force'' (at
separations $d$ much smaller than the capillary length) for which $\bF_{\rm
  bulk}$ is due to gravity.
On the contrary, $Q_0$ will vanish if the system is mechanically
isolated (so that any bulk forces acting on the particle are, according to the
action--reaction principle, equal --- but of oppositte sign --- to any bulk
forces acting on the interface and hence $\bF_{\rm bulk}={\bf 0}$).
This was the key point of a recent controversy as the force in
\eq{eq:Flog} was advocated to explain certain experimental
observations \cite{NBHD02,DKB04,SCLN04} while missing that mechanical
isolation (as purported in the experiments) rules out this force
\cite{MeAi03,FoWu04,ODD05a,ODD06,ODTD07}.

If $Q_0=0$, the capillary force is determined by higher-order terms in
the multipole expansion~(\ref{eq:multipole}). Mechanical isolation
implies the vanishing of the net bulk force and torque, i.e., ``capillary
monopole'' and ``capillary dipole'' (see Eqs.~(\ref{eq:mono_general},
\ref{eq:dipole})). Thus in general, $\bF_\parallel$ will take the form of
a force between quadrupoles, i.e., it is anisotropic and scales as
\begin{equation}
  \label{eq:Fquad}
  |\bF_\parallel| \propto \frac{|Q_2|^2}{d^5}.
\end{equation}
An experimental realization of this case corresponds to nonspherical
inert particles, so that $\Pi\equiv 0$ and the
interfacial deformation is solely due to an undulated contact line (for
a recent corresponding experimental study see \rcite{LYP06}). Concerning
the possibility to relate corresponding experimental observations
and theoretical descriptions we point out the
difficulty that $Q_2$ and higher ``capillary poles'' are
known only in terms of the deformation field $u$, in contrast to $Q_0$
and $Q_1$, which are given by the directly measurable and independently accessible 
quantities ``bulk force'' and ``bulk torque'', respectively.

Another experimentally relevant situation of mechanical isolation
corresponds to the case that the ``capillary charges'' $Q_s$ of all
orders $s\geq 0$ vanish. For example, for an electrically charged,
spherical particle in mechanical isolation one has $Q_0=0$ and $Q_1=0$ by mechanical isolation, 
and by symmetry a rotationally invariant interfacial deformation in
the single--particle configuration, giving $Q_s=0$ also for any
$s\geq 2$. In this case \eq{eq:multipole} reduces to the correction
$\Delta u_R$ and the computation of $\bF_\parallel$ requires a specific model for
$\Pi(\brp)$ and a detailed calculation. In view of our present purposes, we
qualitatively derive only a bound on how rapidly $\bF_\parallel$
decays as function of the separation $d$.

The correction $\Delta u_R(r_R\to\infty)$ has a contribution $\sim
(r_R)^{2-n}$ already in the single--particle configuration because the
rotationally symmetric ``charge density'' $\Pi(r_R\to\infty) \sim
(r_R)^{-n}$ does not have a compact support (see
Appendix~\ref{app:multipolar}).  This provides a contribution to the
capillary force $\bF_\parallel$ which is equal to the force between the
``capillary charges'' at each side of but near the midline because 
the presence of the second particle breaks the rotational symmetry.
This force can be estimated to decay like $d^{-1+2(2-n)}$
because the net ``capillary charge'' in the region farther than a
distance $d$ from one particle is $\sim \int_{r_\parallel > d}
dA_\parallel \; \Pi \sim d^{2-n}$ and the force between charges decays
$\sim d^{\mbox{}-1}$.

Additionally, $\Delta u_R$ has genuine two--particle contributions
``induced'' by the second particle which can be modeled by means of
``induced capillary charges'' $Q_s(d)$ depending on the particle
separation. Generically the dominant term will be a ``capillary
dipole''\footnote{The net ``capillary monopole'' of the halfplane must
  vanish exactly due to mechanical isolation and symmetry reasons.
  Although the net torque on the whole system also vanishes, in this case
  symmetry considerations do not exclude that the net torque on
  one halfplane is opposite to the net torque on the opposite
  halfplane, so that a ``capillary dipole'' in one halfplane is
  possible.} giving rise to a correspondingly dipole-dipole force
$\sim |Q_1(d)|^2 /d^3$, where the ``induced dipole'' $Q_1(d)$ must
decay at least like the inducing field. This is caused by various reasons. 
If $Q_1(d)$ arises by a violation of the boundary conditions
at the contact line, one has $Q_1(d) = {\cal O}(\Delta u_R(d)) \sim
d^{2-n}$, and the force decays by a factor $1/d^2$ faster than the
contribution discussed above concerning the violation of rotational
symmetry.
But there can also be deviations from the linear superposition of the pressure on
the interface. This occurs, e.g., if the interfacial deformation is
due to electric fields emanating from the particles, so that $\Pi
\propto E^2$ and $\Pi(\br_\parallel) - [\Pi_{\rm (single)}(\br_R) +
\Pi_{\rm (single)}(\br_L)] \propto E_{\rm (single)}(\br_R) \, E_{\rm
  (single)}(\br_L)$. In this case one has $Q_1(d) = {\cal O}(E_{\rm
  (single)}(d)) \sim d^{-n/2}$.
Thus we can conclude that the lateral force must decay as function of separation
asymptotically at least as
\begin{equation}
  \label{eq:Fmu}
  F_\parallel = {\cal O}(d^{\mbox{}-\min{(2n-3,n+3)}}) ,
\end{equation}
depending on the value of $n$. In any case this force decays more
rapidly than the expression given by \eq{eq:Flog}.

It is instructive to compare $F_\parallel$ with the force obtained
within the energy approach. The latter consists of finding the parametric
dependence on $d$ of the free energy for the two--particle
configuration,
\begin{equation}
  \label{eq:Vmen}
  V_\mathrm{men} = \int_{{\cal S}_\parallel} \upd A_\parallel \; \left[ 
    \, \frac{\gamma}{2} \, |\nablap u|^2 - \Pi \, u \right] 
  + V_\mathrm{part} ,
\end{equation}
where the integral is the contribution by the interface and
$V_\mathrm{part}$ collects the direct contribution by the particles
\cite{ODD05a,DOD06a}. (Within the electrostatic analogy, the effect of
$V_\mathrm{part}$ would be replaced by appropriate boundary conditions
on the ``potential'' $u$ at the interface--particle contact lines.)
$V_\mathrm{men}(d)$ plays the role of a ``potential of mean force''
for the particle--particle interaction, giving rise to a corresponding ``mean
force''
\begin{equation}
  \label{eq:Fmen}
  F_\mathrm{men}(d) = \mbox{} - V'_\mathrm{men}(d) 
\end{equation}
upon integrating out the capillary degrees of freedom within thermal equilibrium. 
One should keep in mind that this approach captures only the mean--fieldlike contribution to the mean force. The capillary wavelike
fluctuations of the interface around the mean meniscus profile
generates additional, Casimir--like contributions to the force
\cite{LOD06,LeOe07}, which we do not consider in the following.

One can distinguish two cases. First, if there are ``permanent
capillary charges'' ($Q_0\neq 0$ if the system is not mechanically
isolated, or $Q_s\neq 0$ for some $s\geq 2$ for mechanical isolation,
as discussed above), $F_\parallel$ coincides with $F_\mathrm{men}$;
see, e.g., \rcites{Nico49,CHW81,ODD05a} for a derivation of
\eq{eq:Flog} or \rcites{SDJ00,KDD01,FoGa02,VSH05} for obtaining
\eq{eq:Fquad}
in the context of the energy approach with the simplifying assumption
that the interface deformation is small
everywhere\footnote{Reference~\onlinecite{KPDI93} performs an exhaustive comparison
  of the two approaches for the special case that gravity is the only
  source of deformation 
  and the interfacial deformation is small everywhere.}. The reasoning
presented here extends, however, this result also to the case that the
deformation around the particles is not small, requiring this only
near the midline between the particles, i.e., asymptotically for
$d\to\infty$.
Furthermore, the electrostatic analogy shows immediately 
that the capillary
forces $F_\parallel$ are asymptotically pairwise additive in
a configuration with more than two particles provided they
possess a nonvanishing ``permanent capillary pole''.

The second case corresponds to the absence of 
``permanent capillary charges'' as described above. This has been thoroughly
investigated in \rcites{ODD05b,WuFo05,DOD06a} within the energy
approach, and has led to $F_\mathrm{men}(d) \sim d^{-1-n/2}$, which does
not agree with any of the possible asymptotic decays indicated in
\eq{eq:Fmu}.
In order to understand this discrepancy, we recall that by
definition (\eq{eq:Flateral}) $\bF_\parallel$ 
represents the net force acting on the subsystem formed by the
particle {\em and} the piece of interface enclosed by the contour
indicated in Fig.~\ref{fig:2coll}. The work done by this force upon an
infinitesimal virtual displacement $\delta d$ is not related in any
simple manner to the change $\delta V_\mathrm{men}$, which according
to the definition in \eq{eq:Vmen} will involve the work done by local
forces during the rearrangement of the ``capillary charges'' {\em
  inside} the subsystem, so that in general $F_\parallel \neq F_\mathrm{men}$. 

In conclusion one has $F_\parallel = F_\mathrm{men}$ if the $d$--dependence of
$V_\mathrm{men}(d)$ is dominated asymptotically by a multipole
expansion, i.e.,
the whole subsystem can be replaced by a set of point ``capillary
poles'': the degrees of freedom related to the internal
structure 
are irrelevant and only the separation $d$ and the orientation of the
``capillary poles'' matter.
This is related to the validity of the ``superposition
approximation''\cite{Nico49} usually employed in the energy approach, which
consists of approximating the deformation field $u$ by the sum of the
deformation fields induced by each particle in the single--particle
configuration. In \rcite{DOD06a} it is shown that this
approximation is valid if the system is not mechanically isolated
because in that case, asymptotically for $d\to\infty$, the
interface--mediated effect of one particle on the other amounts to
shift it --- together with its surrounding interface ---
vertically as a whole, i.e., without probing or affecting the
``internal structure'' of the subsystem ``particle plus surrounding
interface''.

Finally, it is clear that both $F_\parallel$, defined in
\eq{eq:Flateral}, and $F_\mathrm{men}$, defined in \eq{eq:Fmen},
include a contribution from $\Pi$
and thus differ from the force acting {\em only} on the colloidal
particle (which would be the integral in \eq{eq:Flateral} extended only along the
particle--interface contact line). 
If one is interested in physical situations in thermal equilibrium,
$F_\mathrm{men}$ does represent the effective force between the
particles, i.e., once the capillary degrees of freedom have been
integrated out. 
In dynamical situations out of equilibrium, completely new
considerations have to be made concerning, e.g., whether the capillary
degrees of freedom can be assumed to have relaxed towards thermal
equilibrium in the dynamical time scale of interest. But this
discussion lies beyond the scope of the present analysis.

\section{Nonplanar reference interface}
\label{sec:nonflat}

In the following we shall discuss some applications of \eq{eq:equilibrium}
for particles trapped at an interface which in its
unperturbed state is curved. 
In Subsec.~\ref{sec:droplet} we shall first consider the interfacial
deformation induced by a single charged particle on an otherwise
spherical droplet.
This configuration is particularly relevant for the experiment
described in \rcite{NBHD02}. As explained in the previous section,
mechanical isolation rules out a monopolarlike (i.e., logarithmic)
deformation if the unperturbed interface is flat. Since there has
been recently a controversy whether this conclusion is altered by the
curvature of the droplet, we shall present a thorough analysis for such systems.
In Subsec.~\ref{sec:curved} the application of the electrostatic
analogy to a generally curved interface is illustrated.

\subsection{Particle on a spherical droplet}
\label{sec:droplet}

\begin{figure}
  \centering{\includegraphics[width=.65\textwidth]{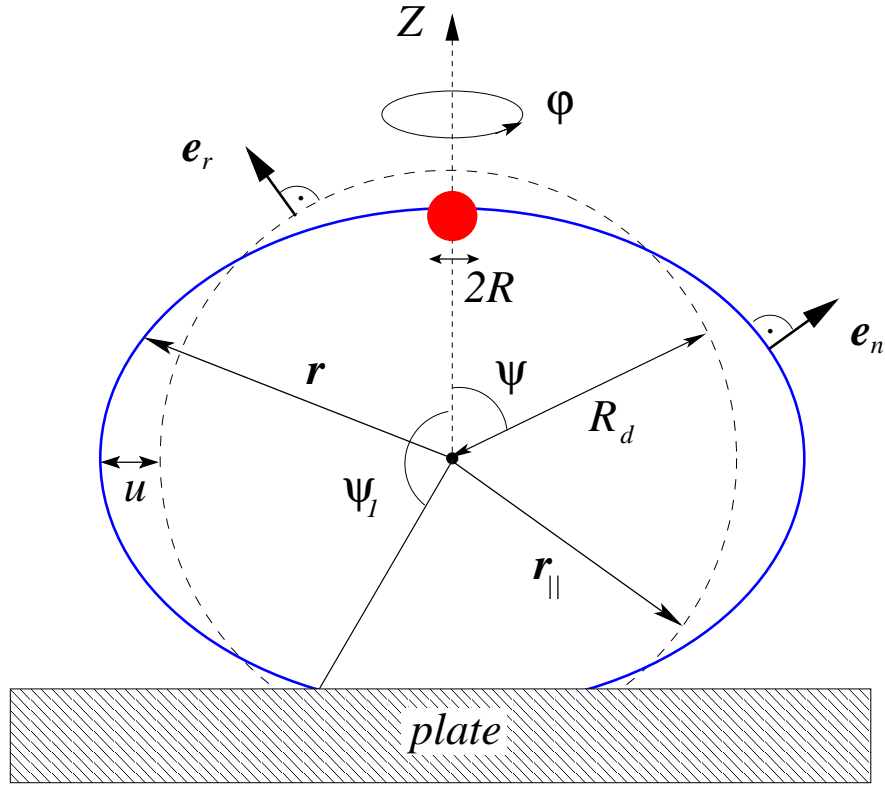}}
  \caption{Charged colloidal particle (radius $R$) at the interface of
    a droplet residing on a plate. Positions on the interface are
    parametrized by the polar angle $\psi$ and the revolution angle
    $\varphi$.
    Without the colloidal particle and neglecting gravity
    the droplet has spherical shape
    with radius $\rdrop$ and normal vector $\er$. $\brp = \rdrop
    \er(\psi,\varphi)$ is a point at the unperturbed, spherical
    interface (dashed line) whereas $\br = \brp + u(\psi,\varphi) \er$ is a point at
    the perturbed, nonspherical interface (full line). The perturbed interface
    intersects the plate for $\psi=\psi_1$.}
    \label{fig:droplet}
\end{figure}

We consider a charged spherical 
particle of radius $R$ trapped at the interface of a droplet which
resides on a plate (Fig.~\ref{fig:droplet}). This configuration models
the experiment described in \rcite{NBHD02} in the absence of
  gravity.
Our goal is to compute the deformation of the droplet far from the
particle. Compared with the energy approach, the force approach has
two advantages:
(i) The result is more general because we have to assume only that the
deformation is small {\em far} from the particle; the usual linear
approximation is not required to hold also near the particle. (ii)
The boundary condition ``mechanical equilibrium of the particle'' is
incorporated easily 
irrespective of the details how the particle is attached to the
interface. It will turn out that the implementation of this condition
has been the source of mistakes in the literature.

We apply \eq{eq:equilibrium} to the piece ${\cal S}(\Psi)$ of the
curved interface bounded on one side by the particle--interface contact
line ${\cal C}_0$
and on the other side by a circle given by the constant latitude
$\Psi$, ${\cal C}(\Psi) := \{\psi=\Psi \leq \psi_1 \}$, so that
$\partial{\cal S} = {\cal C}_0 \cup {\cal C}(\Psi)$, and we assume that
the particle is located at the apex opposite to the plate
(Fig.~\ref{fig:droplet}).  The unperturbed state corresponds to an
uncharged particle which does not exert a force on the interface, so
that the equilibrium shape of the interface is spherical. In the
presence of electric charges, the interface will deform. If the
particle stays at the upper apex, also the deformed interface exhibits
axial symmetry.
We rewrite \eq{eq:equilibrium} in three steps. \\

\begin{figure}
  \centering{\includegraphics[width=.65\textwidth]{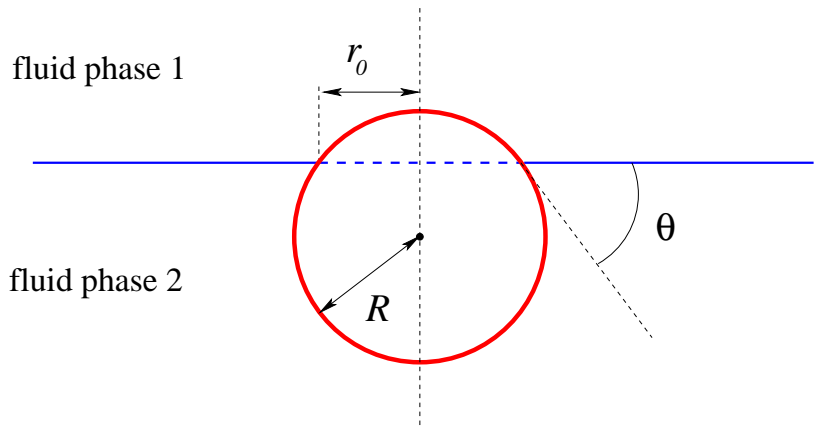}}
  \caption{Side view of a single electrically neutral and spherical
    particle at a planar fluid interface in equilibrium. The radius
    $r_0 = R\sin\theta$ of the circular contact line follows from the
    equilibrium contact angle $\theta$. If the interface was that of
    a large droplet (radius $\rdrop\gg R$, see
    Fig.~\ref{fig:droplet}), this expression would exhibit correction
    terms of order $R/\rdrop$. In the same manner, the presence of
    charges would deform the interface and introduce corrections which
    in first order are proportional to the deformation.}
  \label{fig:ref}
\end{figure}

\noindent
(i) The pressure splits into 
\begin{equation}
  \label{eq:splitPi}
  \Pi (\br) = \Delta p + \Pi_\mathrm{el}(\br) , \qquad 
  \Delta p = \frac{2\gamma}{\rdrop} \left( 1- \mu/2\right) .
\end{equation}
Here, $\rdrop$ is the radius of the unperturbed, spherical droplet,
and $2\gamma/\rdrop$ is the pressure jump across the interface in the
unperturbed state. The dimensionless constant $\mu$ accounts for the
change in hydrostatic pressure due to enforcing the condition of
constant droplet volume in the presence of interface deformations.
$\Pi_\mathrm{el}(\br)$ is the pressure field created by the electric
field emanating from the particle which includes electric stresses and
an osmotic pressure due to a possible discontinuity of the ion
concentrations at the interface (see, e.g.,
\rcites{SMN02,Wuer06a}); this pressure field follows from solving the
corresponding electrostatic problem. We write
\begin{equation}
  \int_{{\cal S}} \upd A \; \en \Pi_\mathrm{el} = 
  2\pi\gamma r_0 \, \vepsp -
  \int_{\cal S_\mathrm{men}\backslash \cal S} \!\!\!\!\!\! \upd A \; 
  \en \Pi_\mathrm{el} ,
\end{equation}
where we have introduced the dimensionless electric force acting on
the whole interface ${\cal S}_\mathrm{men}$ (which at the apex has a
hole carved out),
\begin{equation}
  \label{eq:epspi}
  \vepsp := \frac{1}{2\pi\gamma r_0}
  \int_{\cal S_\mathrm{men}} \!\!\!\!\!\! \upd A \; 
  \en \Pi_\mathrm{el} .
\end{equation}
Here 
$r_0=R\sin\theta$,
where $\theta$ is the equilibrium contact angle between the fluid
phases and the particle; $r_0$ is actually the radius of the circular contact
line of the uncharged particle at a planar interface
(see Fig.~\ref{fig:ref}). \\

\noindent
(ii) Since the particle is in mechanical equilibrium, the
contact line force must be balanced by the hydrostatic force
$\bF_{\Delta p}$ and the electric force $\bF_\mathrm{el}$ acting on
the particle, as well as by any other force $\bF_\mathrm{ext}$ of
external origin (e.g., an optical tweezer pushing or pulling the particle):
\begin{equation}
  \label{eq:contact}
  \gamma \oint_{{\cal C}_0} \upd\ell \; \et \times \en = 
  \bF_{\Delta p} - 2\pi\gamma r_0 (\vepsf + \vepse) ,
\end{equation}
where we have introduced the dimensionless electric and external
forces acting on the particle:
\begin{equation}
  \label{eq:epsf}
  \vepsf := - \frac{\bF_\mathrm{el}}{2\pi\gamma r_0} , \quad
  \vepse := - \frac{\bF_\mathrm{ext}}{2\pi\gamma r_0} .    
\end{equation}
\\

\noindent
(iii) Finally, the following identity is a consequence of elementary
considerations of hydrostatics because $\Delta p$ in \eq{eq:splitPi} is
spatially constant:
\begin{equation}
  \label{eq:hydro}
  \int_{{\cal S}(\Psi)} \upd A \; \en \Delta p + \bF_{\Delta p} = 
  \ez A_z(\Psi) \Delta p , 
\end{equation}
where $A_z(\Psi)$ is the circular area in the $XY$ plane bounded by 
the circle ${\cal C}(\Psi)$.

Thus \eq{eq:equilibrium} can be rewritten as
\begin{equation}
  \label{eq:Spsi}
  \gamma \oint_{{\cal C}(\Psi)} \upd\ell \; \et \times \en = 
  2\pi\gamma r_0 (\vepse+\vepsf - \vepsp) 
  - \ez A_z(\Psi) \Delta p
  + \int_{\cal S_\mathrm{men}\backslash\cal S} \!\!\!\!\!\! \upd A \; 
  \en \Pi_\mathrm{el} . 
\end{equation}
With the notation introduced in Fig.~\ref{fig:droplet}, one has (as in
Sec.~\ref{sec:stress}, the subscript $_\parallel$ denotes quantities
evaluated at and operators acting in the undeformed spherical
interface, i.e., in tangent planes of the undeformed interface):
\begin{subequations}
  \label{eq:quasispher}
\begin{eqnarray}
  \en \, \upd A & = & 
  \left(\frac{\partial \br}{\partial \psi} \times 
    \frac{\partial \br}{\partial \varphi} \right) 
  \upd\psi \upd\varphi = 
  \left( 1 + \frac{u}{\rdrop} \right) 
  \left[\left( 1 + \frac{u}{\rdrop} \right) \er 
    - \nablap u \right] \upd A_\parallel , \\
  & & \nonumber \\
  \et \, \upd\ell & = & (\upd\boldsymbol{\ell}_\parallel \cdot \nabla_\parallel) \br = 
  \left[ 1 + \frac{u}{\rdrop} \right] \upd\boldsymbol{\ell}_\parallel +
  \er (\upd\boldsymbol{\ell}_\parallel \cdot \nablap u) ,
\end{eqnarray}
\end{subequations}
where $\upd\boldsymbol{\ell}_\parallel = \rdrop ( {\bf e}_\psi \upd
\psi + {\bf e}_\varphi \sin\psi \, \upd \varphi )$, $\upd A_\parallel
= \rdrop^2 \sin\psi \, \upd\psi \upd\varphi$,
and
\begin{equation}
  \nablap = \frac{{\bf e}_\psi}{\rdrop}
  \frac{\partial}{\partial \psi}
  +\frac{{\bf e}_\varphi}{\rdrop \sin\psi}
  \frac{\partial}{\partial\varphi}.
\end{equation}
With this notation, $A_z(\Psi) = \pi [\rdrop+u(\Psi)]^2  \sin^2\Psi$.
Equation~(\ref{eq:Spsi}) can be simplified under the assumption that deviations of
the actual droplet shape from the spherical one of radius $\rdrop$ are
small in the distant region $R/\rdrop \ll \Psi \leq \psi < \psi_1$, so
that the linearized approximation of the deformation is valid and
terms quadratic in the quantities $u$, $\mu$, and $\Pi_\mathrm{el}$
(which vanish in the unperturbed state) can be omitted \cite{ODD05a}.
We emphasize that this condition does not exclude large deviations
within the piece ${\cal S}(\Psi)$, in particular near the particle.
Due to rotational symmetry, the vectorial \eq{eq:Spsi} is independent
of the angle $\varphi$ and involves only vectors parallel to $\ez$.
One obtains the following ordinary differential equation for the
function $u(\Psi)$:
\begin{equation}
  \label{eq:1stintegral}
  \sin\Psi \cos\Psi \frac{\partial u}{\partial\Psi} = 
  r_0\, \ez\cdot(\vepse+\vepsf-\vepsp) + 
  \left(\frac{1}{2}\mu\rdrop - u\right) \sin^2\Psi +
  \frac{\rdrop^2}{\gamma} \int_{\Psi}^{\psi_1} \!\!\! \upd\psi 
  \; \sin\psi \cos\psi \; 
  \Pi_\mathrm{el}(\psi) .
\end{equation}
This expression is actually the first integral of the 
Young--Laplace equation (Eq.~(B7) in \rcite{ODD05a}) incorporating the
boundary condition at the contact line.
In terms of the functions $P(\psi):=\cos\psi$, $Q(\psi):=1+\cos\psi
\, \ln\tan(\psi/2)$ and
\begin{equation}
  S(\psi) := - \frac{\rdrop^2}{\gamma} 
  \int_{\psi}^{\psi_1} \!\!\! \upd s \, 
  \sin s \; \Pi_\mathrm{el}(s)
  [P(\psi) Q(s) - P(s) Q(\psi)] ,
\end{equation}
the general solution is 
\begin{equation}
  \label{eq:usolution} 
  u(\psi) = A P(\psi) + r_0\, \ez\cdot (\vepse+\vepsf-\vepsp) Q(\psi) +
  \frac{1}{2}\mu\rdrop + S(\psi),
\end{equation}
where $\mu$ and the integration constant $A$ can be determined by the
boundary condition at the plate, i.e., $\psi=\psi_1$, and the
incompressibility condition of the droplet \cite{ODD05a}. In the limit
$\rdrop\to\infty$ at fixed $r= \rdrop \psi$, one recovers the results
of an unperturbed, planar interface \cite{DOD05}.
We are particularly interested in the deformation given by
\eq{eq:usolution} in the intermediate range $R/\rdrop \ll \psi \ll 1$,
i.e., far from the particle and from the plate. In this range,
$P(\psi) \sim 1$, $Q(\psi) \sim \ln \psi +\mbox{}$const., whereas
$\Pi_\mathrm{el}(\psi)$ will decay in general as $\psi^{-n}$, implying\footnote{One finds $n=6$ in a realistic
model assuming that the whole system has no net charge \cite{Wuer06a}.
In the presence of a net charge,
there is a monopolar electric field 
far from the particle. This interesting case is beyond the scope of
the present analysis, the conclusions of which only hold if
$\Pi_\mathrm{el}(\psi)$ decays sufficiently fast.}
$S(\psi) \sim \psi^{2-n}$ if $n>2$. Thus due to the
behavior of $Q(\psi)$ there may be a logarithmically varying
asymptotic deformation with the amplitude given by $r_0\, \ez\cdot
(\vepse+\vepsf-\vepsp)$.

\begin{figure}
  \centering{\includegraphics[width=.65\textwidth]{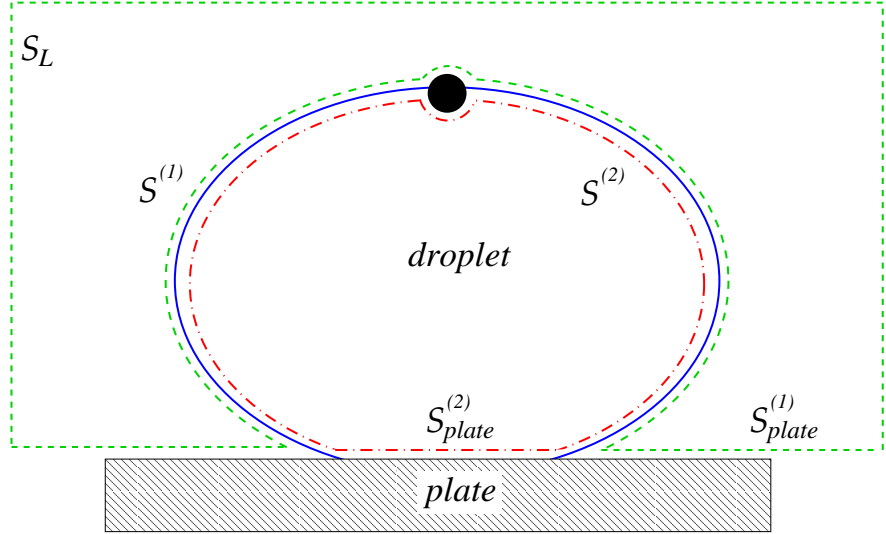}}
  \caption{The surface ${\cal S}^{(1)}$~(${\cal S}^{(2)}$) runs along
    the fluid interface (full line) and the particle (dot) such that
    it lies in the fluid phase exterior~(interior) to the droplet.  The
    surface ${\cal S}_{plate}^{(1)}$~(${\cal S}_{plate}^{(2)}$) is that
    part of the plate surface which is in contact with the fluid phase
    exterior~(interior) to the droplet.  The surface ${\cal S}_L$
    encloses the whole system ``particle + fluid phases'' at a
    macroscopic distance $L\to\infty$ from the droplet. The surfaces
    are oriented towards the exterior of the corresponding fluid volume
    which they enclose.}
  \label{fig:forces}
\end{figure}

We now consider the situation that there is no external field acting
on the system. In this case $\vepse={\bf 0}$ and the electric forces
$\vepsf$ and $\vepsp$ are only due to the charge of the particle. The
value of $\vepsf-\vepsp$ can be obtained by adapting the reasoning of
\rcite{ODD05a} to the droplet geometry. The stress tensor in the fluid
phase exterior to the droplet is given as $\mbox{}-p_1 \mathsf{1} +
\stress_{\rm el}^{(1)}$, where $p_1 \mathsf{1}$ is the homogeneous,
isotropic stress tensor far from the droplet and $\stress_{\rm
  el}^{(1)}$ is Maxwell's stress tensor due to the electric field
(modified to include the possible osmotic pressure by mobile charges,
see, e.g., \rcite{SMN02}). In the same manner, $\mbox{}-(p_1 +\Delta
p) \mathsf{1} + \stress_{\rm el}^{(2)}$ is the stress tensor in the
interior of the droplet, where $\Delta p$ is given by \eq{eq:splitPi}.
With the notations introduced in Fig.~\ref{fig:forces}, the following
equations hold:
\begin{subequations}
  \label{eq:noforce}
  \begin{equation}
    \int_{
      {\cal S}_{plate}^{(1)}\cup{\cal S}_L\cup{\cal S}^{(1)}
    } \upd {\bf A} \cdot \stress_{\rm el}^{(1)} = {\bf 0} ,
  \end{equation}
  \begin{equation}
    \int_{{\cal S}_{plate}^{(2)}\cup{\cal S}^{(2)}} 
    \upd {\bf A} \cdot \stress_{\rm el}^{(2)} = {\bf 0} ,
  \end{equation}
\end{subequations}
which express that the net force on the exterior fluid phase and on
the interior one, respectively, vanishes in equilibrium. (The
contribution to the integrals from the constant isotropic pressures
$p_1$ and $p_1+\Delta p$ is zero.) On the other hand, by definition
one has
\begin{equation}
  2\pi\gamma r_0 (\vepsp-\vepsf) = 
  \mbox{}- \int_{{\cal S}^{(1)}} \upd {\bf A} \cdot 
  \stress_{\rm el}^{(1)}
  - \int_{{\cal S}^{(2)}} \upd {\bf A} \cdot 
  \stress_{\rm el}^{(2)} .
\end{equation}
Combining this with Eqs.~(\ref{eq:noforce}) leads to
\begin{equation}
  \label{eq:nonisolation}
  \vepsp-\vepsf = \frac{1}{2\pi\gamma r_0} \left[
    \int_{{\cal S}_{plate}^{(1)}} \upd {\bf A} \cdot 
    \stress_{\rm el}^{(1)} + 
    \int_{{\cal S}_{plate}^{(2)}} \upd {\bf A} \cdot 
    \stress_{\rm el}^{(2)} 
  \right] ,
\end{equation}
where we have taken into account that the contribution of
$\stress_{\rm el}^{(1)}$ over the surface ${\cal S}_L$ vanishes in the
limit $L\to\infty$ because the electric field decays to zero far away
from the droplet (i.e., we do not consider the possibility that there are external electric fields). 
That is, $\mbox{}-(\vepsp-\vepsf)$ is actually the
(dimensionless) electric force acting on the plate.
The calculation of this integral requires to solve the corresponding
electrostatic problem. However, on dimensional grounds one can obtain
the estimate\footnote{This estimate is supported by explicit
  calculations of $\Pi_\mathrm{el} = \en\cdot(\stress_{\rm
    el}^{(1)}-\stress_{\rm el}^{(2)})\cdot\en$ for realistic
  models \cite{Wuer06a,DaKr06a}. More precisely, the electric force
  exerted on the interface is actually concentrated in a small region
  of area $\sim r_0^2$ around the particle, so that one expects
  $\stress_{\rm el} \sim (\gamma |\vepsp|/r_0) F(r/r_0)$, where $F$ is
  a dimensionless function of order unity at distances $r \sim r_0$
  from the particle and decaying $\sim (r_0/r)^n$ at distances $r \sim
  \rdrop \gg r_0$. On this basis, \eq{eq:nonisolation} provides the
  quoted estimate.}  $|\vepsp-\vepsf| \sim |\vepsp|
(r_0/\rdrop)^{n-2}$. For an asymptotically planar interface, the
logarithmically varying deformation due to nonzero values of
$|\vepsp-\vepsf|$ leads to a long--ranged effective attraction (see
Sec.~\ref{sec:force}), which is the reason that this mechanism has
been invoked to be responsible for the apparent attraction reported in
\rcite{NBHD02}.
If this was the explanation, the measurements in this experiment
would imply a value \cite{ODD05a} $|\vepsp-\vepsf| \sim 10^{-3}$. On
the other hand, the theoretical estimate yields $|\vepsp-\vepsf| \sim
10^{-6}|\vepsp|$ with $n=6$, so that the experimental results would
require $|\vepsp| \sim 10^{3}$. This large value is unlikely for
realistic surface charge densities \cite{ODD05a,ODD05b}.

This result corrects the suggestion made in \rcite{ODD05a} that
$|\vepsp-\vepsf| \sim (r_0/\rdrop)^{2}$, inferred from a not
applicable force balance condition. Indeed, if the deformation is
small also at the contact line, the condition ``mechanical equilibrium
of the particle'' can be derived from \eq{eq:1stintegral} and with an
expansion in terms of $\psi_0 = r_0/\rdrop\ll 1$ leads to
\begin{equation}
  \left.\frac{\upd u}{\upd\psi}\right|_{\psi=\psi_0} \approx 
  \rdrop \, \ez\cdot(\vepse+\vepsf) +
  \psi_0 \left[
    \frac{1}{2} \mu \rdrop - u(\psi_0) \right] .
\end{equation}
The second term, which is subdominant in the limit $\rdrop\to\infty$,
is missing in Eq.~(B8) of \rcite{ODD05a}. We have cross-checked this
corrected expression by deriving it also within the energy approach
employed in \rcite{ODD05a}, which turns out to be algebraically much
more cumbersome.

In conclusion, there persists a logarithmically varying deformation
with an amplitude which is very small in the limit $\rdrop\to\infty$;
this is actually a finite--size effect intrinsic to the geometry of
the set-up and absent for an unbounded flat interface. However, it has the same
physical origin as any logarithmically varying deformation of a flat
interface,
namely that the system ``particle + fluid interface'' cannot be
mechanically isolated in the configuration of a droplet residing on a
solid plate. In the absence of the plate one has $\vepsp-\vepsf={\bf
  0}$ due to \eq{eq:nonisolation} because ${\cal S}_{plate}^{(1)}$ and
${\cal S}_{plate}^{(2)}$ are not there and the logarithmic dependence
in the range $\psi \ll 1$ disappears. This conclusion corrects a
recent claim of the opposite in \rcite{Wuer06b}; the relevant errors
of this work are pinpointed in \rcite{DOD07a}, in particular the
implementation of the boundary condition ``mechanical equilibrium of
the particle''. (Reference~\onlinecite{DOD07a} represents incidentally, within the
energy approach, a further confirmation of our conclusion above.)
To facilitate the comparison of our calculations with \rcite{Wuer06b}
we make two remarks: (i) The reasoning and results are independent of
the precise functional form of the electric pressure
$\Pi_\mathrm{el}(\br)$; the considerations in \rcite{Wuer06b} in this
respect are thus irrelevant. (ii) As a boundary condition for fixing
the droplet \rcite{Wuer06b} employs, instead of a plate at
$\psi=\psi_1$ as used here, a fictional pressure field
$\Pi_\mathrm{com}(\psi) \propto \cos\psi$ constraining the center of
mass of the droplet \cite{MoWi93}. One can easily check that our
general solution (\eq{eq:usolution}) includes this special case, as
the contribution of $\Pi_\mathrm{com}(\psi)$ in \eq{eq:usolution}
eliminates the singularity of $Q(\psi)$ at $\psi=\pi$ and the solution
reduces to the corresponding expression in \rcite{Wuer06b}.  Thus none
of these two issues affects the conclusion concerning the logarithmic
dependence.

\subsection{Particle on a generally curved interface}
\label{sec:curved}

If the particle is trapped at a generally curved interface ({\it
  reference interface}), the electrostatic analogy can still be
exploited provided there is a clear separation of length scales between the
typical radius of curvature $\rdrop$ of the reference interface and
the size of the region around the particle where the interfacial
deformations are appreciable, say, roughly a few times the particle
size $R$. Then, at distances from the particle smaller than $\rdrop$
one can exploit the electrostatic analogy in order to study the small
deviations of the interface from a {\it reference plane}
tangent to the reference interface at some fixed point near the
particle.

The deviations from the reference plane are given by the displacement field
$u(\brp) = u_{\rm ref}(\brp) + \delta u(\brp)$, where $u_{\rm
  ref}(\brp)$ is the deformation of the reference interface and
$\delta u(\brp)$ is the additional deformation brought about by the
presence of the particle. Correspondingly, the pressure field can be
written as $\Pi = \Pi_{\rm ref} + \delta \Pi$.
If $\ell$ is a distance from the particle beyond which the linearized
theory holds (i.e., the deformation near the particle need not be
small), then in the annulus $\ell<r_\parallel<\rdrop$ (with a clear
separation $\ell\ll\rdrop$ so that the following dependence can be
observed) the solution to the field equation can be written as
\begin{eqnarray}
  \label{eq:curvedu}
  \delta u (\brp) & = & B_0 \ln \frac{\zeta}{r_\parallel}
  + \sum_{s=1}^\infty \frac{1}{2 s} \left\{
    \frac{B_s \,{\rm e}^{-i s \varphi} + B_s^* \,{\rm e}^{i s \varphi}}{r_\parallel^s}
    + r_\parallel^s [ A_s \,{\rm e}^{i s \varphi} + A_s^* \,{\rm e}^{-i s \varphi}]
  \right\} \nonumber \\
  & & + \frac{1}{2\pi\gamma} \int_{\ell< r' <\rdrop} dA'_\parallel \;
  \delta\Pi (\brp') \ln\frac{\zeta}{|\brp-\brp'|} .
\end{eqnarray}
The constants $A_s$ and $B_s$ are determined by the boundary conditions
at $r=\ell$ and $r=\rdrop$ and account for ``virtual capillary
charges'' outside the annulus. In particular, we have already seen
that $2\pi\gamma B_0$ is the net bulk force on the region $r<\ell$ in
the direction normal to the reference plane (and in addition to
the net force in the reference state $u_{\rm ref}(\brp)$).

Two new issues arise which we have not considered so far: there is
an ``external boundary'' given by the upper bound $r_\parallel=\rdrop$
and there is a nonvanishing reference deformation $u_{\rm ref}(\brp)$.
The first issue is not exclusive for a curved reference interface and
emerges if the interfacial pressure $\Pi(\brp)$ does not decay
sufficiently fast with the separation from the particle, so that in
principle one cannot carry out the limit $\rdrop\to\infty$. An example has been
studied in full detail in the previous subsection, where $\delta\Pi$
includes a term $-\mu\gamma/\rdrop$ (see \eq{eq:splitPi}) generated by
a nonlocal constraint (i.e., constant volume of the droplet). In general,
the relevance of the boundary conditions at $r_\parallel=\rdrop$
introduces a nonlocal ingredient preventing general
statements based just on the electrostatic analogy localized around
the particle.

The second issue implies an ``external electric field'',
$-\nabla_\parallel u_{\rm ref}$, giving rise to a new phenomenology.
To illustrate this point, we consider a reference minimal surface,
$\nabla_\parallel^2 u_{\rm ref} = 0$ (so that $\Pi_{\rm ref}\equiv
0$), containing a mechanically isolated nonspherical inert particle
(so that $\delta\Pi\equiv 0$) with extension $R$ much smaller than the
typical curvature radius $\rdrop$ of the reference interface.
As we have seen, $Q_0=Q_1=0$ (i.e., no bulk force normal to the
reference plane and no torque in that plane), but $Q_2\neq 0$, so that
the particle will experience a force and a torque which are given, to
leading order in the small ratio $R/\rdrop$, by the coupling of this
quadrupole with the ``external field'' $u_{\rm ref}$. According to
\eq{eq:realcharge}, the real-valued quadrupole is characterized by the
following 2nd rank tensor:
\begin{equation}
  \label{eq:realQ2}
  \hat{Q}_2 = q_2 \left[ (\ex \ex - \ey \ey) \cos 2\theta + 
    (\ex \ey + \ey \ex ) \sin 2\theta \right] ,
\end{equation}
where $q_2 > 0$ is the amplitude and $\theta$ is the angle which the
principal axes of the quadrupole form with the coordinate axes. The
origin of coordinates, $\brp={\bf 0}$, is the contact point of the
reference tangent plane with the interface and must be taken at some
point in or close to the particle, e.g., its center of mass. In these
circumstances, the electrostatic analogy provides the force and the
torque, respectively, as
\begin{equation}
  \bF = \frac{1}{2} \nabla_\parallel \left[ \hat{Q}_2 : 
    \nabla_\parallel \nabla_\parallel u_{\rm ref} \right]_{\brp = {\bf 0}} ,
\end{equation}
\begin{equation}
  {\bf M} = - \nabla_\parallel \times \left[ \hat{Q}_2 \cdot \nabla_\parallel u_{\rm ref} 
  \right]_{\brp = {\bf 0}} ,
\end{equation}
after reversing the sign with respect to the electrostatic
expressions, as discussed above. The most general form of the
traceless Hessian matrix of the reference interface is given by
\begin{equation}
  \label{eq:hessian}
  \nabla_\parallel\nabla_\parallel u_{\rm ref} (\br) = 
  \frac{1}{\rdrop(\br)} 
  \left[ (\ex \ex - \ey \ey) \cos 2\phi(\br) + 
    (\ex \ey + \ey \ex ) \sin 2\phi(\br) \right] ,
\end{equation}
where $R_d(\br)>0$ is the absolute value of the radius of curvature
and $\phi(\br)$ is the angle between the principal directions and the
coordinate axes.
Without loss of generality one can choose the orientation of the
coordinate axes such that $\phi(\brp={\bf 0})=0$. By inserting
Eqs.~(\ref{eq:realQ2}, \ref{eq:hessian}) into the previous expressions
for $\bF$ and ${\bf M}$ one finally obtains
\begin{equation}
  \bF = q_2 \cos 2\theta \, \nabla_\parallel 
  \left(\frac{1}{\rdrop}\right)_{\brp={\bf 0}} +
  2 q_2 \sin 2\theta \, \left. \nabla_\parallel 
    \phi \right|_{\brp={\bf 0}} ,
\end{equation}
\begin{equation}
  {\bf M} = - \frac{2 q_2 \sin 2\theta}{\rdrop(\brp={\bf 0})}\, \ez.
\end{equation}
Therefore, the nonspherical particle tends to rotate so that
$\theta=0,\pi$, i.e., in order to align its ``capillary quadrupole''
with the principal directions of curvature of the reference, minimal interface.
And when aligned like this, it is pulled in the direction of
increasing curvature $1/\rdrop$.
This conclusion
complements the result found in \rcite{Wuer06c}, where the energy
approach has been applied in order to determine the ``potential of
mean force'' of a spherical\footnote{Therefore, the ``capillary
  quadrupole'' is not permanent, as in the present illustrative
  example, but rather induced by the ``external potential'' $u_{\rm
    ref}(\brp)$ and, according to \rcite{Wuer06c}, is proportional to
  the curvature $1/\rdrop$.} inert particle in a minimal
surface. In a multiparticle configuration, this ``external'' force and
this ``external'' torque compete with the capillary interaction
between the quadrupoles (see \eq{eq:Fquad}), possibly leading to
interesting phenomena concerning the 2D patterns formed by the
particles.

\section{Summary and outlook}
\label{sec:end}

We have investigated the force approach for describing colloidal particles
trapped at a fluid interface.
This approach has allowed us to derive a stress--tensor formulation of
the interface--mediated elastic forces for an arbitrary pressure field
$\Pi(\br)$ acting on the interface.  In this manner we have been able
to generalize some of the results of \rcite{MDG05a} obtained only for
a spatially constant pressure field. It is an interesting, open
question whether this result is extendible to, e.g., membranes, for
which bending rigidity as well as surface tension are relevant, and to
other, more general cases considered in \rcite{MDG05a}
and involving constitutive parameters beyond surface tension and
bending rigidity.
Based on the stress--tensor formulation we
have worked out a detailed analogy between small interfacial
deformations and 2D electrostatics, encompassing not only the field
equation of the deformation but also the elastic forces transmitted by
the interface.

We have exploited the electrostatic analogy in order to compute the dominant
contribution to the interface--mediated force between two particles if
they are far apart. This analogy enabled us to clarify the
relationship between the energy and the force approach and to reveal the
advantages and limitations of each. The definition of the effective
force $F_\parallel$, which we have employed in the force approach,
differs from the effective force $F_\mathrm{men}$ introduced in the
energy approach via a ``potential of mean force''. However, the
difference is asymptotically negligible if the interfacial
deformation in the single--particle configuration is expressible as a
multipole expansion (i.e., via nonvanishing ``capillary poles'').
For example, if the system is not mechanically isolated or the
particles are nonspherical, the force approach allows one to
extend with relative ease the energy--approach result to cases in which
the interfacial deformations are not small everywhere. Moreover, it justifies
asymptotic pairwise additivity of the force in a multiparticle
configuration.
One must
bear in mind that none of the two definitions $F_\parallel$ and $F_\mathrm{men}$ 
is the actual force
acting {\em only} on the colloid, because both take into account the
force acting on the surrounding interface. This matters for discussing
the dynamics of trapped particles. However, in thermal equilibrium
$F_\mathrm{men}$ is the effective force according to which the
equilibrium state of the particles is determined. In this situation,
the energy approach, which provides $F_\mathrm{men}$, is in principle
advantageous, while the force approach is more powerful (in the sense that
it may facilitate or extend the range of validity of the calculations) 
whenever it can be shown that $F_\parallel \approx F_\mathrm{men}$.

For the experimentally interesting system of a particle at
the interface of a spherical droplet in contact with a plate
(Fig.~\ref{fig:droplet}), we find that the presence of the plate
breaks mechanical isolation and leads to a logarithmically varying
interfacial deformation at distances $r$ from the particle in the
intermediate range ``particle radius $\ll r \ll$ droplet radius'',
with the amplitude of the logarithm vanishing as the droplet radius
tends to infinity. Our approach has put this finite--size effect on a
sound basis. Nevertheless, our numerical estimates show that 
this logarithmically varying deformation is very likely too weak
in order to explain the apparently long--ranged attraction observed
experimentally in \rcite{NBHD02} for such a system.
However, there are still open questions which we have not addressed
but which are conceivably relevant for this experiment. We have
assumed an electrically neutral system; but if there is a net charge,
e.g., if the colloidal particle is charged but the droplet is not
grounded, additional, long--ranged electric fields arise. Another
interesting question is the loss of rotational symmetry which occurs
if the particle is not fixed at the apex of the droplet: this might
give rise to an additional force (electrostatic or capillary) pushing
the particle towards the apex which, in a multiparticle configuration,
could be misinterpreted as an effective attraction like the one 
apparently observed also for a planar interface. 

Finally, we have discussed briefly the application of the
electrostatic analogy and the associated phenomena arising when the
unperturbed interface is curved. As an illustrative example, we showed
that a nonspherical inert particle trapped at a minimal surface is
pulled to regions in the interface with larger curvature.

\appendix

\section{Torque balance}
\label{app:dipole}

If a piece of interface ${\cal S}$ is in equilibrium, the total torque
on this piece must vanish. In this case, the following condition must
hold (using the same notation as in \eq{eq:equilibrium}):
\begin{equation}
  \label{eq:torque}
  \int_{\cal S} \upd A \; (\br \times \en)\, \Pi + 
  \gamma \oint_{\partial{\cal S}} \upd\ell \; \br \times 
  (\et \times \en) = {\bf 0} .
\end{equation}
If the deviations from a flat interface are small, one can simplify
\eq{eq:torque} as in Sec.~\ref{sec:stress} for the force--balance
equation. To lowest order in the deformation one obtains
\begin{equation}
  \label{eq:torque_lin}
  \left\{ 
    \int_{{\cal S}_\parallel} \upd A_\parallel \; \brp \Pi + 
    \gamma \oint_{\partial{\cal S}_\parallel} \upd\ell_\parallel \; 
    [ \brp (\bn \cdot \nablap u) - u\bn ]
  \right\} \times \ez = {\bf 0} .
\end{equation}
This equation implies that the expression in curly brackets vanishes
because it is a vector orthogonal to $\ez$
($\brp$ and $\bn$ lie in the $XY$ plane).
In the electrostatic analogy the integral over $\Pi$ corresponds to
the ``capillary dipole'' ${\bf P}$ of the piece ${\cal S}$. This
allows one to rewrite \eq{eq:torque_lin} as
\begin{equation}
  {\bf P} = \mbox{} - 
  \gamma \oint_{\partial{\cal S}_\parallel} \upd\ell_\parallel \; 
  [ \brp (\bn \cdot \nablap u) - u\bn ] ,
\end{equation}
which generalizes Gauss' theorem (\eq{eq:gauss}) by expressing the
dipole of a region only in terms of the values of the deformation field and its
derivatives at the boundary. On the other hand, via the general
equilibrium condition in \eq{eq:torque}, the right hand side of
this equality is related to the torque ${\bf M}_{\rm bulk}$ due to the
bulk force:
\begin{equation}
  \gamma \oint_{\partial{\cal S}_\parallel} \upd\ell_\parallel \; 
  [ \brp (\bn \cdot \nablap u) - u\bn ] = 
  \mbox{} - \ez \times {\bf M}_{\rm bulk} .
\end{equation}
The validity of this expression only requires that the deformation is
small at the contour $\partial {\cal S}$, where the linearization is
performed, but not inside. This proofs \eq{eq:dipole}.

\section{Multipole expansion in 2D}
\label{app:multipolar}

Here we recall briefly some results concerning the multipole
expansion in two dimensions. The ``potential'' $u(\br)$ created by a
``charge'' distribution $\Pi(\br)$ is given by ($\zeta$ is an
arbitrary constant)
\begin{equation}
  \label{eq:2dfield}
  u(\br) = -\frac{1}{2\pi\gamma} \int \upd A'\, \Pi(\br') 
  \ln\frac{|\br-\br'|}{\zeta} =
  -\frac{1}{2\pi\gamma} Re \int \upd A'\, \Pi(\br') 
  \ln\frac{z-z'}{\zeta} ;
\end{equation}
the second equation introduces the complex variable $z=r
\exp{(i\varphi)}$ in order to ease the calculations with $Re$ denoting
the real part. We first consider the case that $\Pi$ has a compact
support: $\Pi(\br)=0$ if $r>R$. The Taylor expansion
\begin{equation}
  \label{eq:logTaylor}
  \ln(z-z') = \ln z - \sum_{s=1}^\infty 
  \frac{1}{s} \left(\frac{z'}{z}\right)^s
\end{equation}
is valid in the complex domain $|z'| < |z|$. Inserting this expansion
into the general expression~(\ref{eq:2dfield}) one obtains
straightforwardly
\begin{equation}
  \label{eq:multexpansion}
  u(\br) = \frac{Q_0}{2\pi\gamma} \ln \frac{\zeta}{r}
  + \frac{1}{2\pi\gamma} \sum_{s=1}^\infty \frac{Q_s \,{\rm e}^{-i s \varphi} + 
    Q_s^* \,{\rm e}^{i s \varphi}}{2 s r^s} ,  
\end{equation}
valid for $r>R$, with the complex-valued multipolar charges $Q_s$ given by
\eq{eq:charge}. As can be easily deduced from this latter expression,
they can be written as $Q_s = q_s \exp{(i s \theta_s)}$, where the
amplitude $q_s$ is a positive real number and $\theta_s \in [0,2\pi)$
is the angle by which the configuration with the charge $q_s$ is to be
rotated in order to achieve a configuration with the charge $Q_s$. By
using the identity $r \exp{(-i\varphi)} = \br\cdot(\ex - i \ey)$, the
expansion~(\ref{eq:multexpansion}) can be rewritten in a more familiar
form involving only real-valued quantities:
\begin{equation}
  u(\br) = \frac{\hat{Q}_0}{2\pi\gamma} \ln \frac{\zeta}{r}
  + \frac{1}{2\pi\gamma} \sum_{s=1}^\infty 
  \frac{\er \stackrel{(s)}{\dots} \er}{s r^s} \bullet \hat{Q}_s,  
\end{equation}
where $\bullet$ indicates $s$ scalar products, $\er=\br/r$, and
\begin{equation}
  \label{eq:realcharge}
  \hat{Q}_s = Re [Q_s (\ex - i \ey) \stackrel{(s)}{\dots} 
  (\ex - i \ey) ] =
  q_s Re [{\rm e}^{i s \theta_s} (\ex - i \ey) \stackrel{(s)}{\dots} 
  (\ex - i \ey) ] 
\end{equation}
are the real-valued multipolar charges.

Assume now that $\Pi\sim r^{-n}$ as $r\to\infty$, so that $Q_s$ is
ill-defined for $s\geq n-2$. Nevertheless, one can still write
\begin{equation}
  u(\br) = \frac{Q_0}{2\pi\gamma} \ln \frac{\zeta}{r}
  + \frac{1}{2\pi\gamma} \sum_{s=1}^\nu 
  \frac{Q_s \,{\rm e}^{-i s \varphi} + Q_s^* \,{\rm e}^{i s \varphi}}{2 s r^s} + 
  \Delta u(\br) ,  
\end{equation}
where $\nu$ is the largest integer such that $\nu<n-2$. This
expression serves to define $\Delta u$.
By using the Taylor expansion~(\ref{eq:logTaylor}) again, one can
write
\begin{eqnarray}
  2\pi\gamma \Delta u & = & Re \int_{|z'|<|z|} \upd A'\, \Pi(\br') \,
  \sum_{s=\nu+1}^\infty \frac{1}{s}\left(\frac{z'}{z}\right)^s \nonumber \\
  & & + Re \int_{|z|<|z'|} \upd A'\, \Pi(\br') \,
  \left[ \ln\frac{z}{z'}
    + \sum_{s=1}^\infty \frac{1}{s} \left(\frac{z}{z'}\right)^s
    - \sum_{s=1}^\nu \frac{1}{s} \left(\frac{z'}{z}\right)^s 
  \right] .
\end{eqnarray}
In this form one can easily check that $\Delta u \sim r^{2-n}$ 
for $r=|z|\to\infty$, and the finite multipole
expansion~(\ref{eq:multipole}) holds with an extra term which is
asymptotically indeed subdominant.


\begin{thebibliography}{42}
\expandafter\ifx\csname natexlab\endcsname\relax\def\natexlab#1{#1}\fi
\expandafter\ifx\csname bibnamefont\endcsname\relax
  \def\bibnamefont#1{#1}\fi
\expandafter\ifx\csname bibfnamefont\endcsname\relax
  \def\bibfnamefont#1{#1}\fi
\expandafter\ifx\csname citenamefont\endcsname\relax
  \def\citenamefont#1{#1}\fi
\expandafter\ifx\csname url\endcsname\relax
  \def\url#1{\texttt{#1}}\fi
\expandafter\ifx\csname urlprefix\endcsname\relax\def\urlprefix{URL }\fi
\providecommand{\bibinfo}[2]{#2}
\providecommand{\eprint}[2][]{\url{#2}}

\bibitem[{\citenamefont{Ghezzi and Earnshaw}(1997)}]{GhEa97}
\bibinfo{author}{\bibfnamefont{F.}~\bibnamefont{Ghezzi}} \bibnamefont{and}
  \bibinfo{author}{\bibfnamefont{J.}~\bibnamefont{Earnshaw}},
  \bibinfo{journal}{J.\ Phys.: Condensed\ Matt.} \textbf{\bibinfo{volume}{9}},
  \bibinfo{pages}{L517} (\bibinfo{year}{1997}).

\bibitem[{\citenamefont{Ruiz-Garc{\'\i}a
  et~al.}(1997)\citenamefont{Ruiz-Garc{\'\i}a, G\'amez-Corrales, and
  Ivlev}}]{RGI97}
\bibinfo{author}{\bibfnamefont{J.}~\bibnamefont{Ruiz-Garc{\'\i}a}},
  \bibinfo{author}{\bibfnamefont{R.}~\bibnamefont{G\'amez-Corrales}},
  \bibnamefont{and} \bibinfo{author}{\bibfnamefont{B.~I.} \bibnamefont{Ivlev}},
  \bibinfo{journal}{Physica A} \textbf{\bibinfo{volume}{236}},
  \bibinfo{pages}{97} (\bibinfo{year}{1997}).

\bibitem[{\citenamefont{Stamou et~al.}(2000)\citenamefont{Stamou, Duschl, and
  Johannsmann}}]{SDJ00}
\bibinfo{author}{\bibfnamefont{D.}~\bibnamefont{Stamou}},
  \bibinfo{author}{\bibfnamefont{C.}~\bibnamefont{Duschl}}, \bibnamefont{and}
  \bibinfo{author}{\bibfnamefont{D.}~\bibnamefont{Johannsmann}},
  \bibinfo{journal}{\pre} \textbf{\bibinfo{volume}{62}}, \bibinfo{pages}{5263}
  (\bibinfo{year}{2000}).

\bibitem[{\citenamefont{Quesada-P\'erez
  et~al.}(2001)\citenamefont{Quesada-P\'erez, Moncho-Jord\'a,
  Mart{\'\i}nez-L\'opez, and Hidalgo-Alvarez}}]{QMMH01}
\bibinfo{author}{\bibfnamefont{M.}~\bibnamefont{Quesada-P\'erez}},
  \bibinfo{author}{\bibfnamefont{A.}~\bibnamefont{Moncho-Jord\'a}},
  \bibinfo{author}{\bibfnamefont{F.}~\bibnamefont{Mart{\'\i}nez-L\'opez}},
  \bibnamefont{and}
  \bibinfo{author}{\bibfnamefont{R.}~\bibnamefont{Hidalgo-Alvarez}},
  \bibinfo{journal}{J.\ Chem.\ Phys.} \textbf{\bibinfo{volume}{115}},
  \bibinfo{pages}{10897} (\bibinfo{year}{2001}).

\bibitem[{\citenamefont{Nikolaides et~al.}(2002)\citenamefont{Nikolaides,
  Bausch, Hsu, Dinsmore, Brenner, Gay, and Weitz}}]{NBHD02}
\bibinfo{author}{\bibfnamefont{M.~G.} \bibnamefont{Nikolaides}},
  \bibinfo{author}{\bibfnamefont{A.~R.} \bibnamefont{Bausch}},
  \bibinfo{author}{\bibfnamefont{M.~F.} \bibnamefont{Hsu}},
  \bibinfo{author}{\bibfnamefont{A.~D.} \bibnamefont{Dinsmore}},
  \bibinfo{author}{\bibfnamefont{M.~P.} \bibnamefont{Brenner}},
  \bibinfo{author}{\bibfnamefont{C.}~\bibnamefont{Gay}}, \bibnamefont{and}
  \bibinfo{author}{\bibfnamefont{D.~A.} \bibnamefont{Weitz}},
  \bibinfo{journal}{Nature} \textbf{\bibinfo{volume}{420}},
  \bibinfo{pages}{299} (\bibinfo{year}{2002}).

\bibitem[{\citenamefont{Tolnai et~al.}(2003)\citenamefont{Tolnai, Agod,
  Kabai-Faix, Kov{\'a}cs, Ramsden, and H{\'o}rv{\"o}lgyi}}]{TAKK03}
\bibinfo{author}{\bibfnamefont{G.}~\bibnamefont{Tolnai}},
  \bibinfo{author}{\bibfnamefont{A.}~\bibnamefont{Agod}},
  \bibinfo{author}{\bibfnamefont{M.}~\bibnamefont{Kabai-Faix}},
  \bibinfo{author}{\bibfnamefont{A.~L.} \bibnamefont{Kov{\'a}cs}},
  \bibinfo{author}{\bibfnamefont{J.~J.} \bibnamefont{Ramsden}},
  \bibnamefont{and}
  \bibinfo{author}{\bibfnamefont{Z.}~\bibnamefont{H{\'o}rv{\"o}lgyi}},
  \bibinfo{journal}{J. Phys. Chem. B} \textbf{\bibinfo{volume}{107}},
  \bibinfo{pages}{11109} (\bibinfo{year}{2003}).

\bibitem[{\citenamefont{Fern\'andez-Toledano
  et~al.}(2004)\citenamefont{Fern\'andez-Toledano, Moncho-Jord\'a,
  Mart{\'\i}nez-L\'opez, and Hidalgo-Alvarez}}]{FMMH04}
\bibinfo{author}{\bibfnamefont{J.~C.} \bibnamefont{Fern\'andez-Toledano}},
  \bibinfo{author}{\bibfnamefont{A.}~\bibnamefont{Moncho-Jord\'a}},
  \bibinfo{author}{\bibfnamefont{F.}~\bibnamefont{Mart{\'\i}nez-L\'opez}},
  \bibnamefont{and}
  \bibinfo{author}{\bibfnamefont{R.}~\bibnamefont{Hidalgo-Alvarez}},
  \bibinfo{journal}{Langmuir} \textbf{\bibinfo{volume}{20}},
  \bibinfo{pages}{6977 } (\bibinfo{year}{2004}).

\bibitem[{\citenamefont{Chen et~al.}(2005)\citenamefont{Chen, Tan, Ng, Ford,
  and Tong}}]{CTNF05}
\bibinfo{author}{\bibfnamefont{W.}~\bibnamefont{Chen}},
  \bibinfo{author}{\bibfnamefont{S.}~\bibnamefont{Tan}},
  \bibinfo{author}{\bibfnamefont{T.-K.} \bibnamefont{Ng}},
  \bibinfo{author}{\bibfnamefont{W.~T.} \bibnamefont{Ford}}, \bibnamefont{and}
  \bibinfo{author}{\bibfnamefont{P.}~\bibnamefont{Tong}},
  \bibinfo{journal}{\prl} \textbf{\bibinfo{volume}{95}},
  \bibinfo{pages}{218301} (\bibinfo{year}{2005}).

\bibitem[{\citenamefont{Nicolson}(1949)}]{Nico49}
\bibinfo{author}{\bibfnamefont{M.~M.} \bibnamefont{Nicolson}},
  \bibinfo{journal}{Proc.\ Cambridge Philos.\ Soc.}
  \textbf{\bibinfo{volume}{45}}, \bibinfo{pages}{288} (\bibinfo{year}{1949}).

\bibitem[{\citenamefont{Chan et~al.}(1981)\citenamefont{Chan, {Henry Jr.}, and
  White}}]{CHW81}
\bibinfo{author}{\bibfnamefont{D.~Y.~C.} \bibnamefont{Chan}},
  \bibinfo{author}{\bibfnamefont{J.~D.} \bibnamefont{{Henry Jr.}}},
  \bibnamefont{and} \bibinfo{author}{\bibfnamefont{L.~R.} \bibnamefont{White}},
  \bibinfo{journal}{J.\ Coll.\ Interface Sci.} \textbf{\bibinfo{volume}{79}},
  \bibinfo{pages}{410} (\bibinfo{year}{1981}).

\bibitem[{\citenamefont{Megens and Aizenberg}(2003)}]{MeAi03}
\bibinfo{author}{\bibfnamefont{M.}~\bibnamefont{Megens}} \bibnamefont{and}
  \bibinfo{author}{\bibfnamefont{J.}~\bibnamefont{Aizenberg}},
  \bibinfo{journal}{Nature} \textbf{\bibinfo{volume}{424}},
  \bibinfo{pages}{1014} (\bibinfo{year}{2003}).

\bibitem[{\citenamefont{Foret and W\"urger}(2004)}]{FoWu04}
\bibinfo{author}{\bibfnamefont{L.}~\bibnamefont{Foret}} \bibnamefont{and}
  \bibinfo{author}{\bibfnamefont{A.}~\bibnamefont{W\"urger}},
  \bibinfo{journal}{\prl} \textbf{\bibinfo{volume}{92}},
  \bibinfo{pages}{058302} (\bibinfo{year}{2004}).

\bibitem[{\citenamefont{Oettel et~al.}(2005{\natexlab{a}})\citenamefont{Oettel,
  Dom{\'\i}nguez, and Dietrich}}]{ODD05a}
\bibinfo{author}{\bibfnamefont{M.}~\bibnamefont{Oettel}},
  \bibinfo{author}{\bibfnamefont{A.}~\bibnamefont{Dom{\'\i}nguez}},
  \bibnamefont{and} \bibinfo{author}{\bibfnamefont{S.}~\bibnamefont{Dietrich}},
  \bibinfo{journal}{\pre} \textbf{\bibinfo{volume}{71}},
  \bibinfo{pages}{051401} (\bibinfo{year}{2005}{\natexlab{a}}).

\bibitem[{\citenamefont{Oettel et~al.}(2005{\natexlab{b}})\citenamefont{Oettel,
  Dom{\'\i}nguez, and Dietrich}}]{ODD05b}
\bibinfo{author}{\bibfnamefont{M.}~\bibnamefont{Oettel}},
  \bibinfo{author}{\bibfnamefont{A.}~\bibnamefont{Dom{\'\i}nguez}},
  \bibnamefont{and} \bibinfo{author}{\bibfnamefont{S.}~\bibnamefont{Dietrich}},
  \bibinfo{journal}{J.\ Phys.: Condensed\ Matt.} \textbf{\bibinfo{volume}{17}},
  \bibinfo{pages}{L337} (\bibinfo{year}{2005}{\natexlab{b}}).

\bibitem[{\citenamefont{W{\"u}rger and Foret}(2005)}]{WuFo05}
\bibinfo{author}{\bibfnamefont{A.}~\bibnamefont{W{\"u}rger}} \bibnamefont{and}
  \bibinfo{author}{\bibfnamefont{L.}~\bibnamefont{Foret}}, \bibinfo{journal}{J.
  Phys. Chem. B} \textbf{\bibinfo{volume}{109}}, \bibinfo{pages}{16435}
  (\bibinfo{year}{2005}).

\bibitem[{\citenamefont{Dom{\'\i}nguez
  et~al.}(2005)\citenamefont{Dom{\'\i}nguez, Oettel, and Dietrich}}]{DOD05}
\bibinfo{author}{\bibfnamefont{A.}~\bibnamefont{Dom{\'\i}nguez}},
  \bibinfo{author}{\bibfnamefont{M.}~\bibnamefont{Oettel}}, \bibnamefont{and}
  \bibinfo{author}{\bibfnamefont{S.}~\bibnamefont{Dietrich}},
  \bibinfo{journal}{J.\ Phys.: Condensed Matt.} \textbf{\bibinfo{volume}{17}},
  \bibinfo{pages}{S3387} (\bibinfo{year}{2005}).

\bibitem[{\citenamefont{Oettel et~al.}(2006)\citenamefont{Oettel,
  Dom{\'\i}nguez, and Dietrich}}]{ODD06}
\bibinfo{author}{\bibfnamefont{M.}~\bibnamefont{Oettel}},
  \bibinfo{author}{\bibfnamefont{A.}~\bibnamefont{Dom{\'\i}nguez}},
  \bibnamefont{and} \bibinfo{author}{\bibfnamefont{S.}~\bibnamefont{Dietrich}},
  \bibinfo{journal}{Langmuir} \textbf{\bibinfo{volume}{22}},
  \bibinfo{pages}{846} (\bibinfo{year}{2006}).

\bibitem[{\citenamefont{Danov et~al.}(2006)\citenamefont{Danov, Kralchevsky,
  and Boneva}}]{DKB06}
\bibinfo{author}{\bibfnamefont{K.~D.} \bibnamefont{Danov}},
  \bibinfo{author}{\bibfnamefont{P.~A.} \bibnamefont{Kralchevsky}},
  \bibnamefont{and} \bibinfo{author}{\bibfnamefont{M.~P.}
  \bibnamefont{Boneva}}, \bibinfo{journal}{Langmuir}
  \textbf{\bibinfo{volume}{22}}, \bibinfo{pages}{2653} (\bibinfo{year}{2006}).

\bibitem[{\citenamefont{Dom{\'\i}nguez
  et~al.}(2007{\natexlab{a}})\citenamefont{Dom{\'\i}nguez, Oettel, and
  Dietrich}}]{DOD06a}
\bibinfo{author}{\bibfnamefont{A.}~\bibnamefont{Dom{\'\i}nguez}},
  \bibinfo{author}{\bibfnamefont{M.}~\bibnamefont{Oettel}}, \bibnamefont{and}
  \bibinfo{author}{\bibfnamefont{S.}~\bibnamefont{Dietrich}},
  \bibinfo{journal}{J. Chem. Phys.} \textbf{\bibinfo{volume}{127}},
  \bibinfo{pages}{204706} (\bibinfo{year}{2007}{\natexlab{a}}).

\bibitem[{\citenamefont{Segel}(1987)}]{Sege77}
\bibinfo{author}{\bibfnamefont{L.~A.} \bibnamefont{Segel}},
  \emph{\bibinfo{title}{Mathematics applied to continuum mechanics}}
  (\bibinfo{publisher}{Dover, New York}, \bibinfo{year}{1987}).

\bibitem[{\citenamefont{Kralchevsky et~al.}(1993)\citenamefont{Kralchevsky,
  Paunov, Denkov, Ivanov, and Nagayama}}]{KPDI93}
\bibinfo{author}{\bibfnamefont{P.~A.} \bibnamefont{Kralchevsky}},
  \bibinfo{author}{\bibfnamefont{V.~N.} \bibnamefont{Paunov}},
  \bibinfo{author}{\bibfnamefont{N.~D.} \bibnamefont{Denkov}},
  \bibinfo{author}{\bibfnamefont{I.~B.} \bibnamefont{Ivanov}},
  \bibnamefont{and} \bibinfo{author}{\bibfnamefont{K.}~\bibnamefont{Nagayama}},
  \bibinfo{journal}{J. Coll. Interface Sci.} \textbf{\bibinfo{volume}{155}},
  \bibinfo{pages}{420} (\bibinfo{year}{1993}).

\bibitem[{\citenamefont{M{\"u}ller et~al.}(2005)\citenamefont{M{\"u}ller,
  Deserno, and Guven}}]{MDG05a}
\bibinfo{author}{\bibfnamefont{M.~M.} \bibnamefont{M{\"u}ller}},
  \bibinfo{author}{\bibfnamefont{M.}~\bibnamefont{Deserno}}, \bibnamefont{and}
  \bibinfo{author}{\bibfnamefont{J.}~\bibnamefont{Guven}},
  \bibinfo{journal}{\epl} \textbf{\bibinfo{volume}{69}}, \bibinfo{pages}{482}
  (\bibinfo{year}{2005}).

\bibitem[{\citenamefont{Aveyard et~al.}(2000)\citenamefont{Aveyard, Clint,
  Nees, and Paunov}}]{ACNP00}
\bibinfo{author}{\bibfnamefont{R.}~\bibnamefont{Aveyard}},
  \bibinfo{author}{\bibfnamefont{J.~H.} \bibnamefont{Clint}},
  \bibinfo{author}{\bibfnamefont{D.}~\bibnamefont{Nees}}, \bibnamefont{and}
  \bibinfo{author}{\bibfnamefont{V.~N.} \bibnamefont{Paunov}},
  \bibinfo{journal}{Langmuir} \textbf{\bibinfo{volume}{16}},
  \bibinfo{pages}{1969} (\bibinfo{year}{2000}).

\bibitem[{\citenamefont{Loudet et~al.}(2006)\citenamefont{Loudet, Yodh, and
  Pouligny}}]{LYP06}
\bibinfo{author}{\bibfnamefont{J.~C.} \bibnamefont{Loudet}},
  \bibinfo{author}{\bibfnamefont{A.~G.} \bibnamefont{Yodh}}, \bibnamefont{and}
  \bibinfo{author}{\bibfnamefont{B.}~\bibnamefont{Pouligny}},
  \bibinfo{journal}{Phys. Rev. Lett.} \textbf{\bibinfo{volume}{97}},
  \bibinfo{pages}{018304} (\bibinfo{year}{2006}).

\bibitem[{\citenamefont{Smalyukh et~al.}(2004)\citenamefont{Smalyukh,
  Chernyshuk, Lev, Nych, Ognysta, Nazarenko, and Lavrentovich}}]{SCLN04}
\bibinfo{author}{\bibfnamefont{I.}~\bibnamefont{Smalyukh}},
  \bibinfo{author}{\bibfnamefont{S.}~\bibnamefont{Chernyshuk}},
  \bibinfo{author}{\bibfnamefont{B.}~\bibnamefont{Lev}},
  \bibinfo{author}{\bibfnamefont{A.}~\bibnamefont{Nych}},
  \bibinfo{author}{\bibfnamefont{U.}~\bibnamefont{Ognysta}},
  \bibinfo{author}{\bibfnamefont{V.}~\bibnamefont{Nazarenko}},
  \bibnamefont{and}
  \bibinfo{author}{\bibfnamefont{O.}~\bibnamefont{Lavrentovich}},
  \bibinfo{journal}{\prl} \textbf{\bibinfo{volume}{93}},
  \bibinfo{pages}{117801} (\bibinfo{year}{2004}).

\bibitem[{\citenamefont{Kralchevsky et~al.}(2001)\citenamefont{Kralchevsky,
  Denkov, and Danov}}]{KDD01}
\bibinfo{author}{\bibfnamefont{P.~A.} \bibnamefont{Kralchevsky}},
  \bibinfo{author}{\bibfnamefont{N.~D.} \bibnamefont{Denkov}},
  \bibnamefont{and} \bibinfo{author}{\bibfnamefont{K.~D.} \bibnamefont{Danov}},
  \bibinfo{journal}{Langmuir} \textbf{\bibinfo{volume}{17}},
  \bibinfo{pages}{7694} (\bibinfo{year}{2001}).

\bibitem[{\citenamefont{Paunov}(1998)}]{Paun98}
\bibinfo{author}{\bibfnamefont{V.~N.} \bibnamefont{Paunov}},
  \bibinfo{journal}{Langmuir} \textbf{\bibinfo{volume}{14}},
  \bibinfo{pages}{5088} (\bibinfo{year}{1998}).

\bibitem[{\citenamefont{Kleman and Lavrentovich}(2003)}]{KlLa03}
\bibinfo{author}{\bibfnamefont{M.}~\bibnamefont{Kleman}} \bibnamefont{and}
  \bibinfo{author}{\bibfnamefont{O.~D.} \bibnamefont{Lavrentovich}},
  \emph{\bibinfo{title}{Soft Matter Physics}} (\bibinfo{publisher}{Springer,
  New York}, \bibinfo{year}{2003}).

\bibitem[{\citenamefont{Hurd}(1985)}]{Hurd85}
\bibinfo{author}{\bibfnamefont{A.~J.} \bibnamefont{Hurd}},
  \bibinfo{journal}{J.\ Phys.\ A: Math.\ Gen.} \textbf{\bibinfo{volume}{18}},
  \bibinfo{pages}{L1055} (\bibinfo{year}{1985}).

\bibitem[{\citenamefont{Danov and Kralchevsky}(2006)}]{DaKr06a}
\bibinfo{author}{\bibfnamefont{K.~D.} \bibnamefont{Danov}} \bibnamefont{and}
  \bibinfo{author}{\bibfnamefont{P.~A.} \bibnamefont{Kralchevsky}},
  \bibinfo{journal}{J. Coll. Interface Sci.} \textbf{\bibinfo{volume}{298}},
  \bibinfo{pages}{213} (\bibinfo{year}{2006}).

\bibitem[{\citenamefont{Dom{\'i}nguez et~al.}()\citenamefont{Dom{\'i}nguez,
  Frydel, and Oettel}}]{DFO08}
\bibinfo{author}{\bibfnamefont{A.}~\bibnamefont{Dom{\'i}nguez}},
  \bibinfo{author}{\bibfnamefont{D.}~\bibnamefont{Frydel}}, \bibnamefont{and}
  \bibinfo{author}{\bibfnamefont{M.}~\bibnamefont{Oettel}},
  \bibinfo{journal}{\pre}
  \textbf{\bibinfo{volume}{77}}, \bibinfo{pages}{020401(R)} (\bibinfo{year}{2008}).

\bibitem[{\citenamefont{Oettel et~al.}(2008)\citenamefont{Oettel,
  Dom{\'\i}nguez, Tasinkevych, and Dietrich}}]{ODTD07}
\bibinfo{author}{\bibfnamefont{M.}~\bibnamefont{Oettel}},
  \bibinfo{author}{\bibfnamefont{A.}~\bibnamefont{Dom{\'\i}nguez}},
  \bibinfo{author}{\bibfnamefont{M.}~\bibnamefont{Tasinkevych}},
  \bibnamefont{and} \bibinfo{author}{\bibfnamefont{S.}~\bibnamefont{Dietrich}},
  \bibinfo{journal}{unpublished}  (\bibinfo{year}{2008}).

\bibitem[{\citenamefont{Danov et~al.}(2004)\citenamefont{Danov, Kralchevsky,
  and Boneva}}]{DKB04}
\bibinfo{author}{\bibfnamefont{K.~D.} \bibnamefont{Danov}},
  \bibinfo{author}{\bibfnamefont{P.~A.} \bibnamefont{Kralchevsky}},
  \bibnamefont{and} \bibinfo{author}{\bibfnamefont{M.~P.}
  \bibnamefont{Boneva}}, \bibinfo{journal}{Langmuir}
  \textbf{\bibinfo{volume}{20}}, \bibinfo{pages}{6139} (\bibinfo{year}{2004}).

\bibitem[{\citenamefont{Lehle et~al.}(2006)\citenamefont{Lehle, Oettel, and
  Dietrich}}]{LOD06}
\bibinfo{author}{\bibfnamefont{H.}~\bibnamefont{Lehle}},
  \bibinfo{author}{\bibfnamefont{M.}~\bibnamefont{Oettel}}, \bibnamefont{and}
  \bibinfo{author}{\bibfnamefont{S.}~\bibnamefont{Dietrich}},
  \bibinfo{journal}{\epl} \textbf{\bibinfo{volume}{75}}, \bibinfo{pages}{174}
  (\bibinfo{year}{2006}).

\bibitem[{\citenamefont{Lehle and Oettel}(2007)}]{LeOe07}
\bibinfo{author}{\bibfnamefont{H.}~\bibnamefont{Lehle}} \bibnamefont{and}
  \bibinfo{author}{\bibfnamefont{M.}~\bibnamefont{Oettel}},
  \bibinfo{journal}{\pre} \textbf{\bibinfo{volume}{75}}, \bibinfo{pages}{011602} (\bibinfo{year}{2007}).

\bibitem[{\citenamefont{Fournier and Galatola}(2002)}]{FoGa02}
\bibinfo{author}{\bibfnamefont{J.-B.} \bibnamefont{Fournier}} \bibnamefont{and}
  \bibinfo{author}{\bibfnamefont{P.}~\bibnamefont{Galatola}},
  \bibinfo{journal}{\pre} \textbf{\bibinfo{volume}{65}},
  \bibinfo{pages}{031601} (\bibinfo{year}{2002}).

\bibitem[{\citenamefont{van Nierop et~al.}(2005)\citenamefont{van Nierop,
  Stijnman, and Hilgenfeldt}}]{VSH05}
\bibinfo{author}{\bibfnamefont{E.}~\bibnamefont{van Nierop}},
  \bibinfo{author}{\bibfnamefont{M.~A.} \bibnamefont{Stijnman}},
  \bibnamefont{and}
  \bibinfo{author}{\bibfnamefont{S.}~\bibnamefont{Hilgenfeldt}},
  \bibinfo{journal}{\epl} \textbf{\bibinfo{volume}{72}}, \bibinfo{pages}{671}
  (\bibinfo{year}{2005}).

\bibitem[{\citenamefont{Shestakov et~al.}(2002)\citenamefont{Shestakov,
  Milovich, and Noy}}]{SMN02}
\bibinfo{author}{\bibfnamefont{A.~I.} \bibnamefont{Shestakov}},
  \bibinfo{author}{\bibfnamefont{J.~L.} \bibnamefont{Milovich}},
  \bibnamefont{and} \bibinfo{author}{\bibfnamefont{A.}~\bibnamefont{Noy}},
  \bibinfo{journal}{J. Coll. Interface Sci.} \textbf{\bibinfo{volume}{247}},
  \bibinfo{pages}{62} (\bibinfo{year}{2002}).

\bibitem[{\citenamefont{W{\"u}rger}(2006{\natexlab{a}})}]{Wuer06a}
\bibinfo{author}{\bibfnamefont{A.}~\bibnamefont{W{\"u}rger}},
  \bibinfo{journal}{Eur. J. Phys. E} \textbf{\bibinfo{volume}{19}},
  \bibinfo{pages}{5} (\bibinfo{year}{2006}{\natexlab{a}}).

\bibitem[{\citenamefont{W{\"u}rger}(2006{\natexlab{b}})}]{Wuer06b}
\bibinfo{author}{\bibfnamefont{A.}~\bibnamefont{W{\"u}rger}},
  \bibinfo{journal}{\epl} \textbf{\bibinfo{volume}{75}}, \bibinfo{pages}{978}
  (\bibinfo{year}{2006}{\natexlab{b}}).

\bibitem[{\citenamefont{Dom{\'\i}nguez
  et~al.}(2007{\natexlab{b}})\citenamefont{Dom{\'\i}nguez, Oettel, and
  Dietrich}}]{DOD07a}
\bibinfo{author}{\bibfnamefont{A.}~\bibnamefont{Dom{\'\i}nguez}},
  \bibinfo{author}{\bibfnamefont{M.}~\bibnamefont{Oettel}}, \bibnamefont{and}
  \bibinfo{author}{\bibfnamefont{S.}~\bibnamefont{Dietrich}},
  \bibinfo{journal}{\epl} \textbf{\bibinfo{volume}{77}}, \bibinfo{pages}{68002}
  (\bibinfo{year}{2007}{\natexlab{b}}).

\bibitem[{\citenamefont{Morse and Witten}(1993)}]{MoWi93}
\bibinfo{author}{\bibfnamefont{D.~C.} \bibnamefont{Morse}} \bibnamefont{and}
  \bibinfo{author}{\bibfnamefont{T.~A.} \bibnamefont{Witten}},
  \bibinfo{journal}{\epl} \textbf{\bibinfo{volume}{22}}, \bibinfo{pages}{549}
  (\bibinfo{year}{1993}).

\bibitem[{\citenamefont{W{\"u}rger}(2006{\natexlab{c}})}]{Wuer06c}
\bibinfo{author}{\bibfnamefont{A.}~\bibnamefont{W{\"u}rger}},
  \bibinfo{journal}{\pre} \textbf{\bibinfo{volume}{74}},
  \bibinfo{pages}{041402} (\bibinfo{year}{2006}{\natexlab{c}}).



\end{thebibliography}
\end{document}